\documentclass[aps,prb,reprint,letterpaper,amsmath,amssymb, superscriptaddress]{revtex4-2}

\usepackage[T1]{fontenc}
\usepackage{geometry}
\usepackage{setspace}
\usepackage{newunicodechar}
\newunicodechar{₁}{\textsubscript{1}}
\usepackage[section]{placeins}
\usepackage{graphicx}
\usepackage{float}
\newfloat{scheme}{htbp}{los}
\floatname{scheme}{Scheme}
\floatname{chart}{Chart}
\newfloat{graph}{htbp}{loh}
\usepackage{amsmath}
\usepackage{color}
\usepackage{amssymb}   
\usepackage{soul}
\usepackage[table]{xcolor}
\usepackage[colorlinks]{hyperref}

\begin{document}

\title{Millisecond spin relaxation times of distinct electron and hole subensembles in MA$_x$FA$_{1-x}$PbI$_3$ perovskite crystals}


\author{Rongrong Hu}
\email[]{rongrong.hu@tu-dortmund.de}
\affiliation{Experimentelle Physik 2, Technische Universit\"at Dortmund, 44227 Dortmund, Germany}
\affiliation{School of Science, Shanghai Institute of Technology, 201418 Shanghai, China}
\author{Sergey~R.~Meliakov}
\affiliation{Experimentelle Physik 2, Technische Universit\"at Dortmund, 44227 Dortmund, Germany}
\author{Dmitri~R.~Yakovlev}
\affiliation{Experimentelle Physik 2, Technische Universit\"at Dortmund, 44227 Dortmund, Germany}
\author{Bekir~Turedi}
\affiliation{Laboratory of Inorganic Chemistry, Department of Chemistry and Applied Biosciences,  ETH Z\"{u}rich, CH-8093 Z\"{u}rich, Switzerland}
\affiliation{Laboratory for Thin Films and Photovoltaics, Empa-Swiss Federal Laboratories for Materials Science and Technology, CH-8600 D\"{u}bendorf, Switzerland}
\author{Maksym~V.~Kovalenko}
\affiliation{Laboratory of Inorganic Chemistry, Department of Chemistry and Applied Biosciences,  ETH Z\"{u}rich, CH-8093 Z\"{u}rich, Switzerland}
\affiliation{Laboratory for Thin Films and Photovoltaics, Empa-Swiss Federal Laboratories for Materials Science and Technology, CH-8600 D\"{u}bendorf, Switzerland}
\author{Manfred Bayer}
\affiliation{Experimentelle Physik 2, Technische Universit\"at Dortmund, 44227 Dortmund, Germany}
\affiliation{Research Center FEMS, Technische Universit\"at Dortmund, 44227 Dortmund, Germany}
\author{Vasilii~V.~Belykh}
\email[]{vasilii.belykh@tu-dortmund.de}
\affiliation{Experimentelle Physik 2, Technische Universit\"at Dortmund, 44227 Dortmund, Germany}


\begin{abstract}
The unique combination of outstanding optical quality and attractive spin  properties opens new avenues for optical spin control in hybrid organic-inorganic perovskite semiconductors. Using the optically detected magnetic resonance technique, we study the spins of electrons and holes in mixed-cation MA$_x$FA$_{1-x}$PbI$_3$ single crystals with $x = 0.4$ and 0.8. Multiple distinct spin subensembles with absolute values of $g$-factor spanning from 2.9 to 3.5 for electrons and from 0.5 to 1.6 for holes are resolved, revealing diverse localization environments. We measure the longitudinal spin relaxation times, $T_1$, reaching 2 ms and remaining in the $\mu$s range even for weakly localized carriers at cryogenic temperatures. The magnetic-field dependence of $T_1$ is dominated by the random nuclear (Overhauser) fields with strengths of $\sim 0.5$~mT for electrons and $\sim 5-12$~mT for holes. The corresponding correlation times of the hyperfine field are determined by carrier hopping between shallow localization sites. The temperature dependence of $T_1$ reveals a weak localization potential of the charge carriers and shows a correlation between $T_1$ and the inhomogeneity of the spin ensemble. These results establish mixed-A-site perovskite single crystals as a promising solid-state platform with long-lived spin states for quantum information applications.
\end{abstract}

\maketitle

\section{Introduction}
Hybrid organic–inorganic lead halide perovskites (HOIPs) have rapidly emerged as one of the most intensively studied semiconductor classes in the past decade due to their exceptional optoelectronic properties~\cite{herz2017charge,Jeong2021,he2023single,wei2019halide}. Beyond conventional photophysics, the possibility of optical spin orientation \cite{nestoklon2018optical,kopteva2024highly} and the inverted band structure \cite{becker2018bright} with spins of electron and hole equal to 1/2, make them an ideal platform for exploring spin-dependent phenomena. Compared with their extensively studied optical properties, spin-related studies on HOIPs are still underdeveloped, especially for the mixed-A-site hybrid organic–inorganic perovskite crystals. One of the defining features of perovskite crystals is the presence of photo-generated electrons and holes, spatially separated at different sites~\cite{kirstein2022lead,kirstein2025resonant,kudlacik2024optical,belykh2019coherent}, with distinct spin relaxation behavior and spin-dependent parameters. Typically, two such spin signals corresponding to electrons and holes are observed in these perovskite crystals. However, a recent study of FAPbBr$_3$ crystals surprisingly reported an additional electron spin species with  slightly different $g$-factor, localized in the potential fluctuations induced by crystal imperfections~\cite{kirstein2024coherent}. Therefore, the mechanism of carrier localization and its impact on the spin properties, especially spin relaxation, have remained an open question.
      
Spin relaxation is characterized by time $T_1$, reflecting the decay of the spin polarization along the magnetic field (diagonal element of the density matrix), and by spin coherence time $T_2$, reflecting the phase loss of the spin component rotating about the magnetic field (off-diagonal element of the density matrix). A long longitudinal carrier spin relaxation time $T_1$ is critical for applications in quantum information technologies. Indeed, $T_1$ limits the spin coherence time $T_2$~\cite{siyushev2014coherent,belykh2021stimulated} and enables efficient dynamic nuclear polarization~\cite{kirstein2022lead,w11v-2v4g}. Electron spin relaxation times reaching milliseconds and even seconds were reported only for strongly localized systems with suppressed spin-orbit coupling, see the review in ref.~\cite{stano2022review}. In particular, $T_1$ approaching 1~ms was reported for inorganic perovskite nanocrystals (NCs) CsPb(Cl,Br)$_3$ \cite{belykh2022submillisecond} and Ni\textsuperscript{2+}-doped CsPb(Br$_{1-x}$Cl$_x$)$_3$  \cite{barak2022uncovering} at cryogenic temperatures. In bulk systems, the spin relaxation of charge carriers is usually enhanced by the spin-orbit interaction, activating the Dyakonov-Perel spin relaxation mechanism \cite{dyakonov1972spin} and leading to spin relaxation times of a few nanoseconds. In perovskites, the Dyakonov-Perel spin relaxation mechanism is suppressed as consequence of their unique property of spatial inversion symmetry~\cite{kopteva2024highly,kopteva2025effect,gribakin2026spin}. In this respect, they are close to organic systems, where weak spin-orbit interaction leads to spin lifetimes in the microsecond and even second range at 100–300~K \cite{pramanik2007observation,mena2024room}. This allows one to expect long spin relaxation times of charge carriers even in bulk perovskite systems. 

The so far reported $T_1$ values in bulk perovskite single crystals are ranging from tens to hundreds of nanoseconds, with examples including  $T_1 = 37$~ns in MAPbI$_3$ (MA = methylammonium)~\cite{kirstein2022spin}, $T_1 = 470$~ns in FAPbBr$_3$ (FA = formamidinium)~\cite{kirstein2024coherent}, and $T_1 = 53$~ns in CsPbBr$_3$~\cite{belykh2019coherent}. These relatively short $T_1$ times are primarily related to the limited potential of pump-probe and spin inertia \cite{heisterkamp2015longitudinal} techniques for measuring $\mu$s-long spin dynamics. Furthermore, these techniques cannot assign spin relaxation times $T_1$ to different spin species having different $g$-factors, even to distinguish $T_1$ for electrons and holes. These disadvantages are resolved in the recently developed resonant spin inertia technique based on optically detected magnetic resonance (ODMR) with additional optical spin orientation of charge carriers \cite{belykh2022selective,belykh2026amr}.
      
In this study, a comprehensive investigation of the spin relaxation dynamics in mixed-A-site hybrid MA$_x$FA$_{1-x}$PbI$_3$ perovskite single crystals is conducted by taking advantage of the resonant spin inertia technique \cite{belykh2022selective,belykh2026amr}. We resolve multiple distinct electron and hole spin subensembles, each characterized by a unique $g$-factor having absolute values in the range of $2.9-3.5$ for electrons and $0.5-1.6$ for holes. Such a wide, discrete $g$-factor distribution directly reflects the coexistence of electrons and holes with different localization as well as nuclear environments. Remarkably, we observe a $T_1$ exceeding 2 ms in a hole subensemble at low temperature, nearly two to three orders of magnitude longer than previously reported in hybrid perovskites and in general in bulk semiconductors. Simultaneously, all other carrier spin subensembles show $T_1$ times of at least several microseconds. We also reveal resonances corresponding to quasi-free electrons and holes. By analyzing their widths we extract the effective nuclear Overhauser fields of $\sim 0.4-0.8$~mT acting on the electrons and $\sim 5-12$~mT acting on the holes, consistent with the Pb-dominated hyperfine interaction. Also, by analyzing the magnetic-field dependence of $T_1$, which varies in the microsecond range, we determine the correlation times related the to nuclear fluctuations of about $\sim 0.04-0.4$ $\mu$s for electrons and $\sim 1-15$~$\mu$s for holes, respectively. These times correspond to hopping of carriers in a weak localizing potential. We show that an increase in the temperature from 1.6 to 7~K, leading to the carrier delocalization, results only in a moderate decrease in $T_1$ which remains in the microsecond range. These findings provide crucial insights into the spin dynamics in HOIP single crystals, and establish them as a promising platform for quantum technologies. 
\section{Results}
\subsection{Basic spin properties of MA$_{x}$FA$_{1-x}$PbI$_3$ perovskites: $g$ factors and $T_1$}
\begin{figure*}
    \centering
    \includegraphics[width=0.9\linewidth]{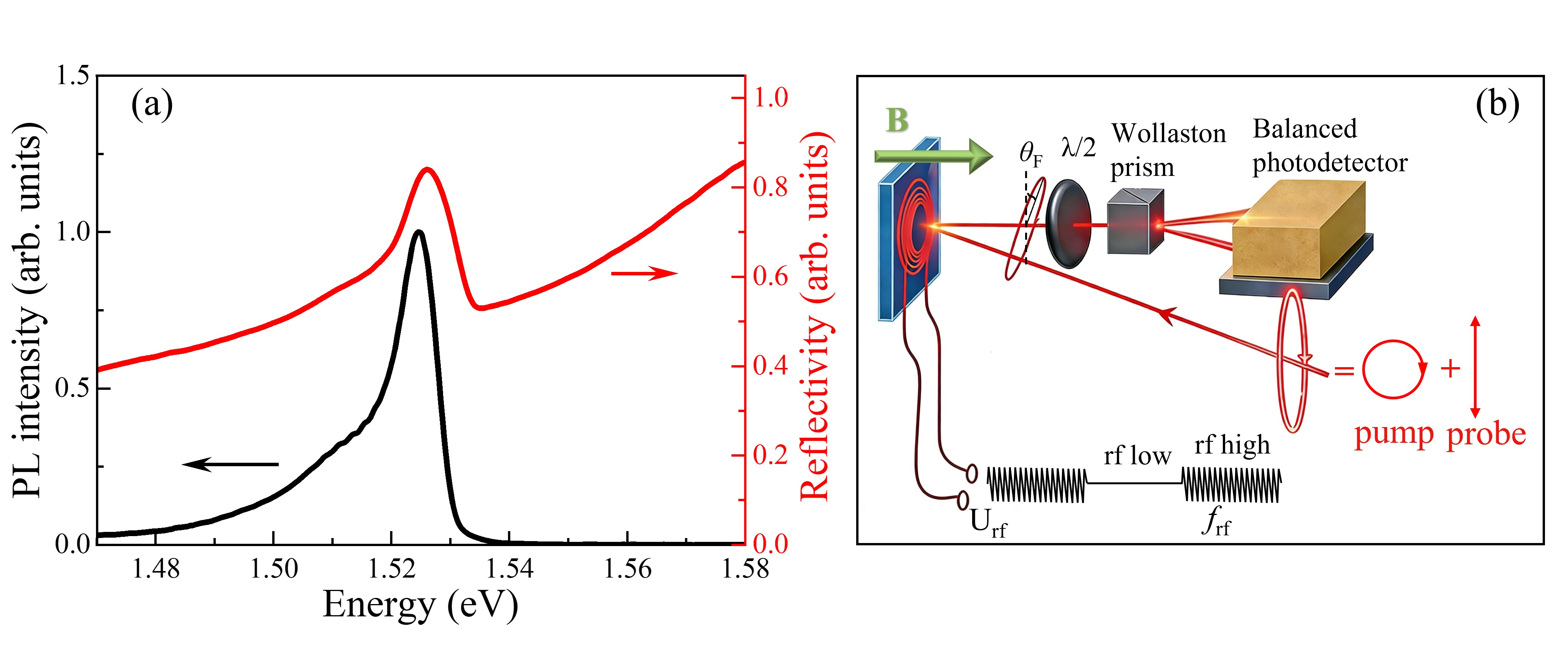}
    \caption{a) Basic optical properties of the MA$_{0.4}$FA$_{0.6}$PbI$_3$ crystal: photoluminescence spectrum (black) measured using continuous-wave excitation with a photon energy of 3.06 eV and reflectivity  spectrum (red) of the MA$_{0.4}$FA$_{0.6}$PbI$_3$ crystal. b) Scheme of the experimental setup for measuring ODMR-based resonant spin inertia.}
    \label{fig:PL, abs and setup}
\end{figure*}

The solution-grown MA$_x$FA$_{1-x}$PbI$_3$ perovskite crystals studied in this work were synthesized using the well-established inverse crystallization  method~\cite{chen2019single,alsalloum2020low}. We investigate two samples with MA content of $x= 0.4$ and 0.8. All experiments are carried out at the temperature of 1.6~K unless specified otherwise. The reflectivity spectrum of the MA$_{0.4}$FA$_{0.6}$PbI$_3$ crystal shown in Figure~\ref{fig:PL, abs and setup}a has the maximum at 1.527~eV, which corresponds to the exciton-polariton resonance. The photoluminescence (PL) spectrum has one line with the maximum at 1.524~eV, with a full width at half maximum of 7 meV, and a shoulder at 1.513~eV. The PL maximum has a slight Stokes shift of 3~meV with respect to the exciton-polariton energy in the reflectivity spectrum. 
  
\begin{figure*}
        \centering
        \includegraphics[width=0.9\linewidth]{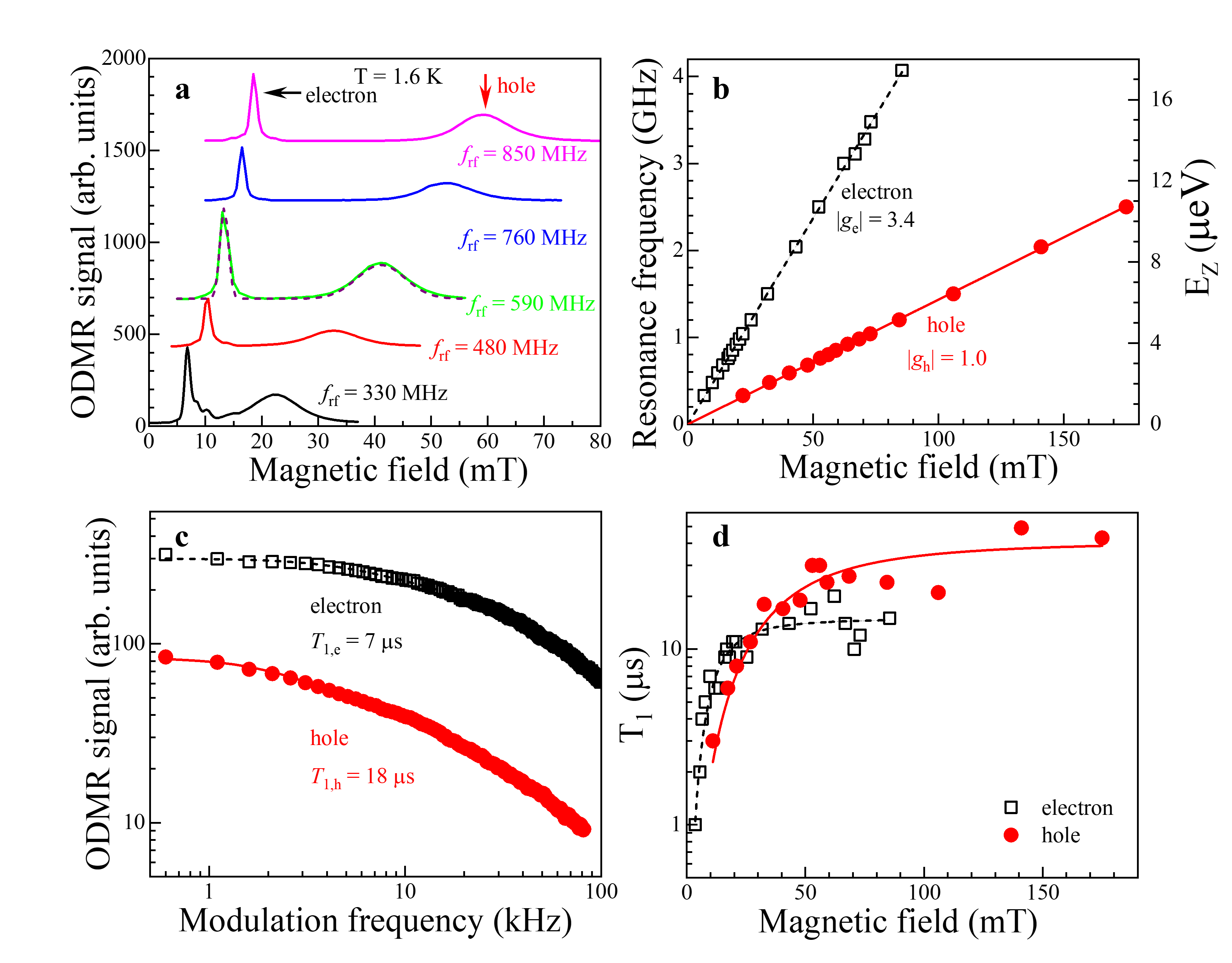}
        \caption{ODMR study of the MA$_{0.4}$FA$_{0.6}$PbI$_3$ crystal. a) ODMR spectra of the MA$_{0.4}$FA$_{0.6}$PbI$_3$ crystal measured at different rf frequencies. The curves are vertically shifted for clarity. The dashed line shows the Gaussian fit. b) Magnetic-field dependence of the resonance frequencies corresponding to electron and hole in the ODMR spectra, with linear fits yielding $|g_\text{e}|=3.4$ and $|g_\text{h}|=1.0$. c) ODMR signal as a function of the modulation frequency, $f_\text{mod}$, at magnetic fields of 9.8 and 32.6~mT corresponding to the electron and hole resonances, respectively, at $f_{\rm rf} = 480$~MHz. The lines show fits of the experimental data with Equations (S1)-(S4). d) Spin relaxation times, $T_1$, for electrons and holes as a function of magnetic field. Lines show fits of the experimental data with Equation~\eqref{eq:T1}. The laser photon energy is 1.528~eV, the laser power is 1~mW. $T=1.6$~K.}
         \label{fig:odmr}
    \end{figure*}
    
To investigate the spin properties of charge carriers in the MA$_x$FA$_{1-x}$PbI$_3$ crystals, the ODMR-based resonant spin inertia technique is employed. Figure~\ref{fig:PL, abs and setup}b shows the experimental setup. The technical details are given in the Experimental section. The carrier spin polarization generated by the circularly polarized component of the laser pulses accumulates along the external magnetic field $\textbf{B}$, which is applied in the Faraday geometry parallel to the sample normal ($\textbf{B} \parallel \textbf{k}$). The spin polarization is monitored via the Kerr rotation of the linearly polarized component. The accumulated spin polarization is destroyed by the rf field, when the spin precession Larmor frequency ($f_{\rm L}$) matches the frequency of the rf field ($f_{\rm rf}$). Applying an rf field with a fixed frequency and scanning the external magnetic field, spin resonances can be recorded in Kerr rotation signals (ODMR signals), as shown in Figure~\ref{fig:odmr}a. ODMR spectra measured in this way show one narrow and one broad peak. Their resonance fields shift with changing the rf field frequency $f_\text{rf}$. The resonant frequency $f_\text{rf} = f_\text{L}$ as a function of the extracted magnetic fields corresponding to the maxima of the two peaks are plotted in Figure~\ref{fig:odmr}b. The dependencies are linear corresponding to 
\begin{equation}
h f_\text{L} = |g| \mu_\text{B} B \,,
\label{eq:wL}
\end{equation}
where $h$ is the Planck constant, and $\mu_\text{B}$ is the Bohr magneton. The slopes of the dependencies in Figure~\ref{fig:odmr}b give the two $g$-factor values of 3.4 and 1.0. The $\textbf{k} \cdot \textbf{p}$ calculations and atomistic modeling \cite{Yu16,Nestoklon21} suggest universal dependencies of the $g$ factor on the bandgap for lead halide perovskites \cite{kirstein2022lande}, which has been confirmed experimentally. According to this dependence, for the bandgap of 1.52~eV, $g_\text{e} > 0$, $g_\text{h} < 0$, and $|g_\text{e}| > |g_\text{h}|~$\cite{Yu16,Nestoklon21}. Therefore, we assign the larger $g_\text{e} = 3.4$ to the electron, while attributing the smaller $g_\text{h}  = -1.0$ to the hole, which is similar to the results obtained from time-resolved optical orientation measurements, i.e. $g_\text{e} = 3.27$ and  $g_\text{h}  = -1.02$ in ref.~\cite{gribakin2026spin}. 

The width of the ODMR peak, $\Delta B$, is determined by the spread of Larmor spin precession frequencies and provides information on the spin dephasing time $T_2^*$. The latter describes the dephasing of the Larmor precession in an inhomogeneous spin ensemble and typically is much shorter than the time $T_2$, which characterizes the coherence of a single spin. We assume a Gaussian distribution of the Larmor frequencies, which is justified by the Gaussian fit in Figure~\ref{fig:odmr}a, with $\Delta B$ being the standard deviation ($\sigma$). In this case the dephasing of the spin precession is described by a Gaussian decay $\propto \exp(-t^2/2T_2^{*2})$~\cite{belykh2016large} with the decay time given by the equation:
\begin{equation}
T_2^* = \frac{\hbar}{|g|\mu_\text{B} \Delta \textit{B}}.
\label{eq:T2*}
\end{equation}
The ODMR peak width $\Delta B$ and, correspondingly, $T_2^*$ are dominated by the random effective fields of the nuclear spin fluctuations and by the spread of the $g$ factors, $\Delta g$~\cite{mikhailov2018electron}. The contribution of the $g$-factor spread to the magnetic field linewidth increases linearly with magnetic field as $\Delta B_g = (\Delta g / g) B$. We do not observe a significant dependence of $T_2^*$ on the magnetic field over the measured range (see Figure S2 in the Supporting Information), suggesting a small contribution of the $g$-factor spread. Thus, $\Delta B$ around 0.5~mT for electrons and around 4~mT for holes correspond to the random effective nuclear fields resulting from the hyperfine interaction. The corresponding spin dephasing times are $T_{2,e}^* =7$~ns for electrons and $T_{2,h}^* =2.8$~ns for holes. These values are in line with the results obtained for FA$_{0.9}$Cs$_{0.1}$PbI$_{2.8}$Br$_{0.2}$ crystals from pump-probe Kerr rotation experiments~\cite{kirstein2022lead}. We note that the hyperfine interaction for holes in perovskite crystals is stronger than that for electrons \cite{kirstein2022lead,Kirstein23}, providing additional proof that the broader ODMR peak corresponds to holes, while the sharper peak corresponds to electrons. 
       
For magnetic field strengths corresponding to the ODMR resonances, we can measure the longitudinal spin relaxation time $T_1$ of electrons and holes using the resonant spin inertia technique~\cite{belykh2022selective,belykh2026amr}. To this end, we modulate the rf field at the frequency $f_\text{mod}$ in the range from 0.1 to 100 kHz and measure ODMR signal as a function of $f_\text{mod}$. Carrier spin polarization accumulates by optical pumping in the half of the period when the rf field is minimal, and then decays by the rf field action during the next half of the period (Figure~\ref{fig:PL, abs and setup}b). The amplitude of the accumulated spin polarization is determined by the carrier spin lifetime $T_1$, when $1/f_\text{mod} \gg T_1$, and by the time $1/f_\text{mod}$, when $1/f_\text{mod} \ll T_1$. Figure~\ref{fig:odmr}c shows the dependence of the ODMR signal (Kerr rotation amplitude) on the modulation frequency of the rf field, $f_\text{mod}$. Increasing the modulation frequency beyond $1/T_1$ leads to a decrease in the ODMR signal for both electrons and holes allowing to estimate $T_1$. More quantitatively, $T_1$ can be determined using the spin inertia equation \cite{belykh2022selective,belykh2022submillisecond,belykh2026amr} for the spin polarization (Kerr rotation amplitude)   
\begin{equation}
S=\frac{A T_1^2}{\sqrt{1+(2\pi T_{1}f_{\mathrm{mod}})^{2}}},
\label{eq:SI} 
\end{equation}
where the parameter $A$ is the frequency-independent coefficient determined by the excitation power and the amplitude of the rf field.
We generalize this equation for a dynamics having two characteristic time scales (see Supporting Information) and use it for fitting the experimental dependencies in Figure~\ref{fig:odmr}c. 
The fits give spin relaxation times of $T_{1,e}=7$~$\mu$s and $T_{1,h}=18$~$\mu$s for electrons and holes, respectively. Note that an increase in the laser power leads to a reduction of $T_1$ through the additional perturbation of the spin system by the laser beam \cite{belykh2022submillisecond}. The dependencies of $1/T_1$ on the laser power $P$ shown in the Supporting Information (Figure S3c) for electrons and holes are linear and their extrapolation to $P = 0$ yields $T_{1,e}= 22$~$\mu$s and $T_{1,h}=88$~$\mu$s, respectively, for the undisturbed spin system. The measured $T_1$ for both electrons and holes are much longer than those of about 100 ns reported so far for other bulk perovskites \cite{kirstein2022lead,kirstein2024coherent,kudlacik2024optical}. 

It is interesting to examine the dependence of $T_1$ for electrons and holes spins on the magnetic field as shown in Figure~\ref{fig:odmr}d. Increase in the magnetic field strength leads to an increase in $T_1$. This is the typical behaviour, as at low fields $B$ the carrier spin dynamics is dominated by the Overhauser field of the nuclear spins varying with the characteristic time $\tau_\text{c}$, the correlation time of the fluctuating nuclear field. The increase in the external magnetic field $B$ leads to suppression of the effect of the nuclear fluctuations and, therefore, $T_1$ increases. The analysis of the magnetic field dependence of $T_1$ allows us to evaluate the correlation time $\tau_\text{c}$ in our experiment. The described scenario was considered in ref. \cite{smirnov2018theory} leading to  the following equation: 
\begin{equation}
 T_{1}(B)=\frac{\tau_\text{s}}{1+(\Delta_\text{N}/B)^{2}(\tau_\text{s}/\tau_\text{c})}.
 \label{eq:T1}
\end{equation}
Here $\Delta_\text{N}$ is the spread of the Overhauser nuclear field distribution $\pi^{-3/2}\Delta_\text{N}^{-3}\exp(-B_\text{N}^2/\Delta_\text{N}^2)$ and $\tau_\text{s}$ is the spin relaxation time in absence of the nuclear fluctuations. This equation is valid for $B$ exceeding $\Delta_\text{N}$ and the corresponding Larmor precession frequency exceeding $1/\tau_\text{c}$. Both conditions are fulfilled in our experiment. 
For the fit we assume values of the Overhauser field of 0.7~mT and 6~mT for electrons and holes, respectively, obtained from the widths of the ODMR peaks $\Delta_\text{N} = \sqrt{2}\Delta B$. From the reasonable fit shown in Figure~\ref{fig:odmr}d, we obtain the values of the effective nuclear correlation times of $\tau_\text{c,e}= 0.04$~$\mu$s for the electrons and $\tau_\text{c,h}= 0.9$~$\mu$s for the holes. 

\begin{figure*}
        \centering
        \includegraphics[width=0.9\linewidth]{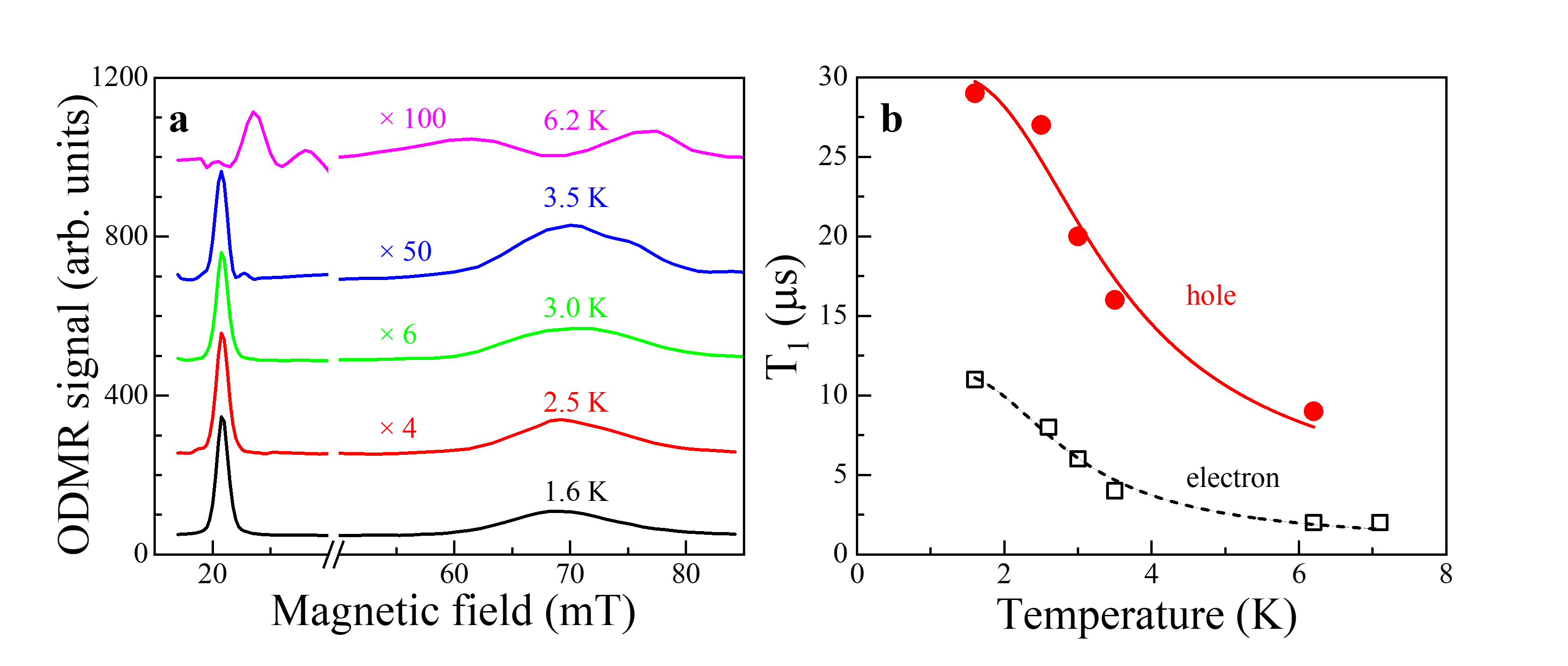}
                \caption{Temperature dependence. a) ODMR spectra measured at different temperatures with the rf frequency fixed at 980~MHz. The curves are vertically shifted and multiplied by the indicated factors for clarity. b) Temperature dependence of the longitudinal spin relaxation time $T_1$ for electron and hole resonances measured at magnetic fields of 20.6 and 68.5~mT, respectively, and $f_\text{rf} = 980$~MHz. The lines are fits using Equation~\eqref{eq:T1T}. For temperatures below 4~K, the laser photon energy is set to 1.528 eV, whereas for temperatures above 4 K, the laser photon energy is 1.530 eV. The laser power is 1 mW. }
        \label{fig:Temperature}
    \end{figure*}

Next we investigate the spin dynamics of carriers at different temperatures ranging from 1.6 to 7.1 K, as shown in Figure~\ref{fig:Temperature}. With increasing temperature, the ODMR signal shows a pronounced reduction in amplitude, see Figure \ref{fig:Temperature}a. $T_1$ decreases with increasing temperature for both electrons and holes, as shown in Figure~\ref{fig:Temperature}b. Note that the $T_1$ for holes remains consistently larger than that for electrons across the investigated temperature range. Remarkably, even at $T = 7.1$~K, the electron time $T_1$ remains as large as 2~$\mu$s. Note that for temperatures above 4~K a slightly higher laser photon energy is used to gain a better signal-to-noise ratio. This change in excitation energy results in a noticeable shift of the electron resonance position at 6.2~K. The temperature dependence of $T_1$ can be well fitted by an activation dependence \cite{kirstein2024coherent}
\begin{equation}
 \frac{1}{T_{1}(T)}=\frac{1}{T_{1}(T=0)}+\gamma_\text{A}\exp\left(-\frac{E_\text{A}}{k_{\mathrm{B}}T}\right) \,.
   \tag{5} 
 \label{eq:T1T}
\end{equation}
Here $\gamma_\text{A}$ is the thermal relaxation rate, $E_\text{A}$ is the activation energy, and $k_\text{B}$ is the Boltzmann constant. The corresponding fits to the experimental data shown by the lines in Figure~\ref{fig:Temperature}b give values of the parameters $E_\text{A,e}=0.86$~meV for electrons and $E_\text{A,h}=0.91$ meV for holes. Such a small energy scale presumably corresponds to shallow potential fluctuations that localize the carriers. In addition, in the Supporting Information (Figure S4), we plot the temperature dependence of the spin dephasing times $T_2^*$ and fit them using the Equation~\eqref{eq:T1T}. The fit yields distinctly different activation energies for electrons and holes: (\textit{E}\textsubscript{A,e} = 3.7 meV and \textit{E}\textsubscript{A,h} = 1.3 meV). The possible origin of these differences is discussed below in the Discussion section.

\subsection{Multiple distinct spin states}

\begin{figure*}
        \centering
        \includegraphics[width=1\linewidth]{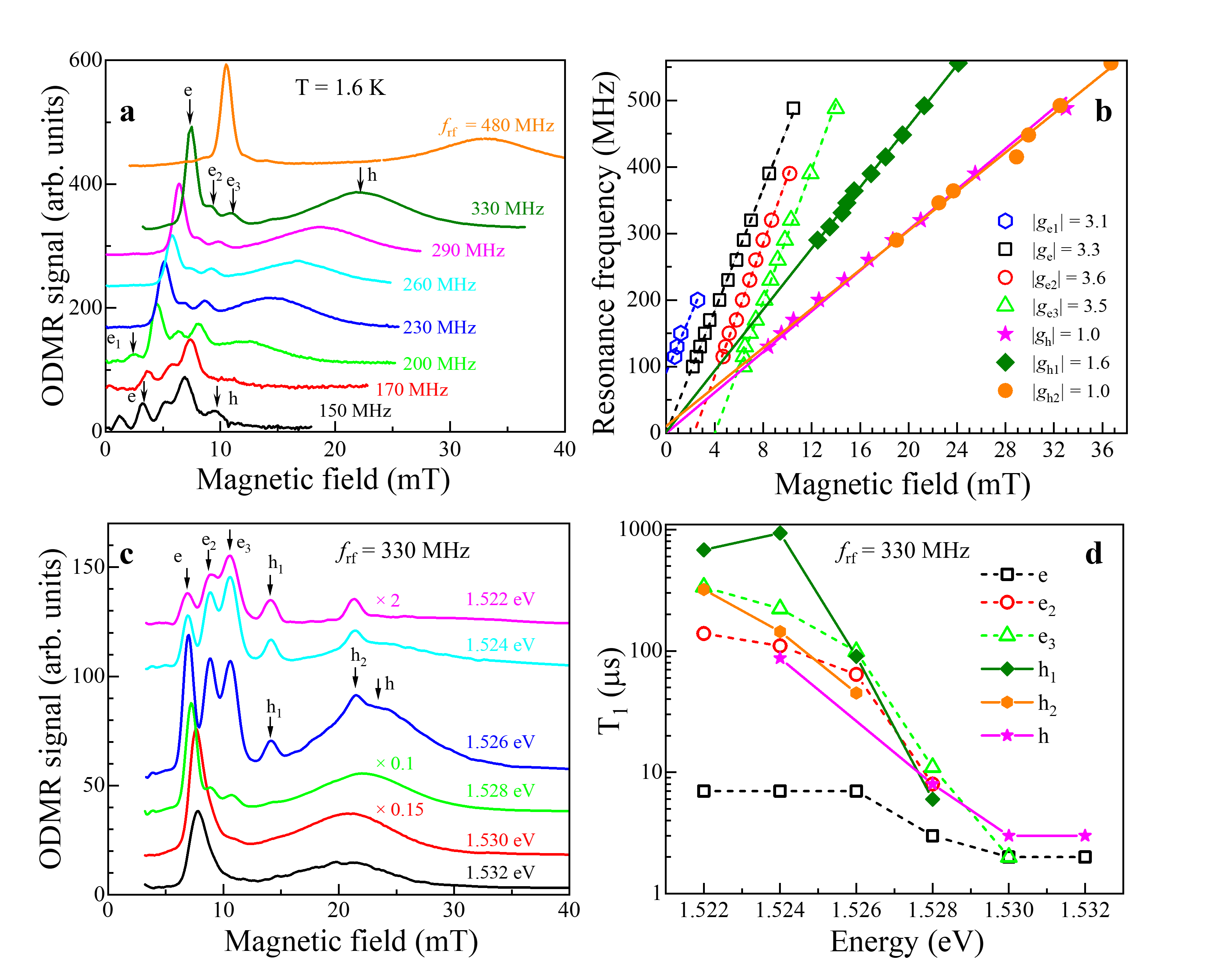}
                \caption{Multiple distinct spin states in the MA$_{0.4}$FA$_{0.6}$PbI$_3$ crystal. a) ODMR spectra of the MA$_{0.4}$FA$_{0.6}$PbI$_3$ crystal measured at different rf frequencies. The curves are vertically shifted for clarity. b) Magnetic field dependence of the resonance frequencies of the ODMR peaks observed in the panel a with linear fits shown by the lines. The laser photon energy is 1.528~eV. The data for h$_1$ and h$_2$ are extracted from the data presented in Figure~S5 with the excitation laser energy of 1.524 eV. c) ODMR spectra at different laser photon energies at fixed \textit{f}\textsubscript{rf} = 330~MHz. d) Spin relaxation times, $T_1$, for different electron and hole resonances as a function of the excitation energy at fixed \textit{f}\textsubscript{rf} = 330~MHz. The laser power is 2 mW, $T= 1.6$~K.}
        \label{fig:fine0.4}
    \end{figure*}

A decrease in $f_{\rm rf}$ down to 150 MHz shifts the electron (labeled as e) and hole (labeled as h) resonances to smaller magnetic fields (Figure~\ref{fig:fine0.4}a). Surprisingly, this also leads to the appearance of new resonances in the ODMR spectrum (labeled as e$_i$ or h$_i$). The frequencies of the all observed resonances that increase linearly with magnetic field (Figure~\ref{fig:fine0.4}b), yield absolute $g$-factor values of 3.1, 3.3, 3.6, 3.5, 1.0, 1.6 and 1.0. Note that the $g$-factor values may slightly vary depending on the excitation energy and the excitation position on the crystal due to inhomogeneity. The four resonances with $g$-factors around 3 likely correspond to distinct spin subensembles, arising from varying strengths of electron localization in potential fluctuations caused by crystal imperfections (e.g., local octahedral distortions), or binding to impurities or point defects \cite{kirstein2024coherent}. Notably, for the additional resonance peaks (e$_1$, e$_2$ and e$_3$), the linear magnetic field dependencies of the resonance frequencies show offsets at zero field. These offsets are unlikely to originate from dynamical nuclear polarization \cite{kirstein2022lead}, because they behave symmetrically upon reversing the magnetic field direction from negative to positive, as demonstrated in Figure~S5b.

Interestingly, the contribution of different resonances to the ODMR spectrum depends not only on the rf frequency, but also on the optical transition energy. In the following, we investigate the evolution of the ODMR spectrum at a fixed rf frequency of \textit{f}\textsubscript{rf} = 330 MHz as a function of the laser photon energy, shown in Figure~\ref{fig:fine0.4}c. Note that the electron spin subensemble with the $g$-factor of 3.1 (marked as e$_1$ in Figure~\ref{fig:fine0.4}b) is not detectable at this rf frequency. At the highest photon energy, the spectrum exhibits only a narrow electron peak (e) and a broad hole peak (h) that were discussed above in the manuscript. A decrease in photon energy leads to the suppression of these peaks and the appearance of other narrow, satellite, peaks. The ODMR signal exhibits a resonant enhancement at the excitation energy of 1.528~eV. Decrease in the laser photon energy to below 1.528 eV leads to the appearance of a new resonant peak (h$_1$)  at 14.7~mT, corresponding to $|g_\text{h1}|= 1.6$, so that it may originate from a hole spin subensemble. Also the broad hole peak becomes accompanied by a more narrow peak (h$_2$) at the field of 21.5 mT with $|g_\text{h2}| = 1.0$, which presumably also has hole origin. Remarkably, the spin dephasing time $T_2^*$ estimated from the ODMR peak width reaches 17 ns for the h$_2$ spin subensemble at the excitation energy of 1.522 eV (see Figure S6 in the Supporting Information), which exceeds previously reported hole spin dephasing times for perovskite materials. Interestingly, the electron peak e slightly shifts towards lower magnetic fields with decreasing laser photon energy in contrast to the hole peak h, which shifts towards higher fields. 
    
Next, we determine the $T_1$ related to the different ODMR resonances as  function of the laser energy (Figure~\ref{fig:fine0.4}d). As the laser energy decreases from 1.532 to 1.522 eV, the time $T_1$ for all spin subensembles significantly increases. Specifically, the spin relaxation time of the electron resonance denoted as e with $|g_\text{e}|= 3.3$ increases only by a factor of 4, from 1.7 $\mu$s to 7 $\mu$s. In contrast, the electron spin subensembles with $|g_\text{e,2}|= 3.6$ (denoted e$_2$) and $|g_\text{e,3}|=3.5$ (denoted e$_3$) exhibit a significantly larger increase in $T_1$, approximately ten-fold and thirty-fold, respectively. For all resonances except of e$_1$, the time $T_1$ can reach several hundreds of $\mu$s (and even 1~ms for h$_1$) at the smallest excitation energy of 1.522 eV. At the laser energy of 1.524 eV, the h$_2$ spin subensemble, corresponding to a narrow ODMR resonance, exhibits a $T_1$ nearly 2 times longer than that for the h spin subensemble showing up as broad peak. The times $T_1$ for the additional resonances also show magnetic field dependence which is presented in Figure~S7 in the Supporting Information.

Similar experimental appearances are found for the MA$_{0.8}$FA$_{0.2}$PbI$_3$ crystal having the MA composition of 0.8 (compared to 0.4), which is discussed in the Supporting Information. These results are shown in Figure~S8 in the Supporting Information. 

\begin{table*}
  \centering
  \begin{tabular}{|c|c|c|c|c|c|c|}
  \hline
 State& $|g|$ & Offset (MHz) & $T_1$ ($\mu$s) &  $\Delta B$ (mT)&$T_2^*$ (ns) &$\tau_\text{c}$ ($\mu$s)\\\hline  \hline

$\mathbf{e}$& $3.3 \pm 0.1$ & $0 \pm 10$ & 7  &  0.5& 7.4& 0.04\\\hline

$\mathbf{e_1}$& $3.1 \pm 0.3$ & $90 \pm 10$ & - &  0.5& 8.0& - \\\hline

$\mathbf{e_2}$& $3.5 \pm 0.2$ & $-110 \pm 20$ & 110 &  0.5& 6.2& - \\\hline

$\mathbf{e_3}$& $3.4 \pm 0.1$ & $-180 \pm 20$ & 220 &  0.7& 4.5& - \\\hline \hline

$\mathbf{h}$& $1.02 \pm 0.05$ &  $0 \pm 20$ & 90 &  4.8 & 2.1& 0.9\\\hline

$\mathbf{h_1}$& $1.61 \pm 0.04$ & $10 \pm 10$ & 940 &  0.6& 17& -  \\\hline

$\mathbf{h_2}$& $1.03 \pm 0.07$ & $10 \pm 30$ & 140 &  0.7& 15& - \\\hline

  \end{tabular}

  \caption{Measured and evaluated parameters of the studied perovskite crystals MA$_{0.4}$FA$_{0.6}$PbI$_3$ at $T= 1.6$~K. The values of $T_1$ and $\Delta B$ are extracted at the laser photon energy of 1.524~eV, with the rf frequency of 330 MHz. Note that the $\Delta B$ of the $e_1$ subensemble is extracted from the data shown in Figure~\ref{fig:fine0.4}a using the rf frequency of 200~MHz.} 

\label{tab:1}
\end{table*}

\begin{table*}
  \centering
  \begin{tabular}{|c|c|c|c|c|c|c|}
  \hline
 State& $|g|$ & Offset (MHz) & $T_1$ ($\mu$s) &  $\Delta B$ (mT)&$T_2^*$ (ns) &$\tau_\text{c}$ ($\mu$s)\\\hline  \hline

$\mathbf{e}$& $2.9 \pm 0.1$& $10 \pm 10$ & 6 & 0.6& 6.0& 0.4\\\hline


$\mathbf{e_2}$&  $3.4 \pm 0.2$ &  $-160 \pm 40$ &  150 & 0.7& 4.8& - \\\hline

$\mathbf{e_3}$& $3.1 \pm 0.1$ &  $-190 \pm 20$ & 230 & 1.1& 3.3& - \\\hline \hline

$\mathbf{h}$& $0.50 \pm 0.05$ &  $10 \pm 10$  & 160 & 11.7& 1.9& 15\\\hline

$\mathbf{h_1}$& $1.2 \pm 0.1$ & $50 \pm 30$ & 2100 & 1.1& 7.3& - \\\hline

$\mathbf{h_2}$& $0.9 \pm 0.1$ & $10 \pm 10$ & 1140 & 0.7& 17& - \\\hline
  \end{tabular}

  \caption{Measured and evaluated parameters of the studied perovskite crystal MA$_{0.8}$FA$_{0.2}$PbI$_3$ at $T= 1.6$~K. The $T_1$ and $\Delta B$ are extracted at the laser photon energy of 1.617 eV, with the rf frequency of 330~MHz.}

\label{tab:2}
\end{table*}

\section{Discussion}

Using the ODMR-based technique we have observed a number of spin resonances for the MA$_x$FA$_{1-x}$PbI$_3$ crystals with $x = 0.4$ and 0.8 and measured the basic spin parameters related to these resonances: $g$ factor, spin precession frequency offset at $B = 0$, longitudinal spin relaxation time $T_1$, and inhomogeneous spin dephasing time $T_2^*$. These spin parameters are summarized in Tables~\ref{tab:1} and~\ref{tab:2} for $x = 0.4$ and 0.8, respectively. Here the uncertainty of $g$ factors and frequency offsets takes into account the spread of these parameters for measurements in different points of the sample and at different laser energies.

Across a wide range of magnetic fields (rf frequencies) and laser energies we observe resonances, denoted in the tables as e and h with absolute $g$ factor values close to 3 and 1, which we attribute to electrons and holes, respectively. They have drastically different widths of about 1 and 10~mT for electrons and holes, respectively, which correspond to the spread of the Overhauser fields of the nuclear spin fluctuations. The hyperfine interaction is much stronger for holes rather than for electrons in perovskites \cite{kirstein2022lead,kotur2026hyperfineinteractionelectronsholes,w11v-2v4g}, which is in line with our experimental results. 

We have measured long spin relaxation times $T_1$ reaching tens of $\mu$s for both electrons and holes, which is surprising for bulk semiconductors. We note that, in general, in our experiments we have measured spin lifetimes which are contributed by both the carrier lifetime and the actual spin relaxation time $T_1$. However, the strong dependence of the measured time on the magnetic field suggests that it is determined mostly by $T_1$ and, thus, we deal with resident carriers having a very long lifetime. The same is true for the additional spin resonances, which also have a strong dependence of the spin lifetime on the magnetic field (see Figure~S7 in the Supporting Information). For the resident carriers the mechanism of optical orientation and spin detection via Faraday rotation is essentially the same as was described in the work \cite{Greilich2006} for singly charged quantum dots.
The times $T_1$ increase with magnetic field (Figures~\ref{fig:odmr}d and S8c) due to the suppression of the time-varying nuclear spin fluctuations. These dependencies allow us to estimate the nuclear field correlation times $\tau_\text{c}$ of 0.04~$\mu$s (0.4~$\mu$s) for electrons and 0.9~$\mu$s (15~$\mu$s) for holes for the samples with $x = 0.4$ ($x = 0.8$). For strongly localized carriers, $\tau_\text{c}$ is determined by the nuclear spin dynamics. However, in lead halide perovskites, the electrons and holes interact with different nuclear species, with the holes primarily coupled to the Pb nuclei and the electrons coupled to both the Pb and I nuclei \cite{kirstein2022lead}. Since the nuclear spin dynamics of Pb and I can differ, a difference in $\tau_\text{c}$ between electrons and holes can be expected even in the strongly localized regime. For weakly localized carriers $\tau_\text{c}$ may be determined by the carrier hopping between potential traps having different nuclear environments, if the time for these hoppings is shorter than the evolution time of the nuclear polarization. In our case, we have strongly different $\tau_\text{c}$ for electrons and holes. Also, $\tau_\text{c}$ is rather different for samples with different MA concentration $x$. These facts suggest that $\tau_\text{c}$ is determined mostly by carrier hopping rather than by nuclear spin dynamics, which is expected to be weakly dependent on carrier type and $x$. This gives us evidence that we deal with carriers weakly localized in shallow potentials, and that the carriers are more delocalized in the sample with reduced MA content. Thus, at low magnetic fields, spin relaxation is dominated by the hyperfine interaction with nuclear spins. At high magnetic fields, spin relaxation might be contributed by the mechanisms related to spin-orbit interaction. As mentioned earlier, inversion symmetry of the perovskite lattice suppresses spin-orbit coupling and, thus, the Dynakonov-Perel spin relaxation mechanism, which is usually very efficient for free carriers \cite{kopteva2024highly,kopteva2025effect,gribakin2026spin}. However, lattice imperfections, residual electric fields, and defects may break this symmetry, especially for localized carriers, which may result in spin-orbit coupling and the corresponding relaxation mechanisms, both of the Dyakonov-Perel type, where spin relaxes as a result of carrier motion and frequent scattering slows down the relaxation, and the Elliott-Yafet type, where spin relaxes as a result of scattering \cite{Meier1984}. Such relaxation may take place for carriers hopping between different localization sites, just as in our case. An additional spin relaxation mechanism, especially important at elevated temperatures, is related to the two-phonon Raman process as described in Ref.~\cite{belykh2026millisecond}.

Another confirmation of a weak localization by shallow potentials comes from the temperature dependence of $T_1$ (Figure~\ref{fig:Temperature}b), which shows an activation behavior with rather small energies of about $1$~meV for both electrons and holes. This energy can be related to the depth of carrier localization potential. We also highlight the decrease in the ODMR signal amplitude when the temperature is increased (Figure~\ref{fig:Temperature}a). Note that the accumulated spin polarization is proportional to $T_1^2$ [Equation~\eqref{eq:SI}]~\cite{belykh2022selective,belykh2026amr}. However, the decrease in $T_1$ cannot fully account for the decrease in the ODMR signal. For example, when the temperature is increased from 1.6 to 3.5~K, the electron $T_\text{1,e}$ decreases from 11 to 6~$\mu$s, corresponding to a three-fold decrease in $T_1^2$, while the experiment reveals a 50-fold decrease in the ODMR signal. This dominant signal suppression can be related to electron and hole delocalization with their subsequent recombination, which, thus, reduces the number of resident carriers that can be oriented optically.

The other observation following from the temperature dependence of the ODMR spectra is the broadening of the ODMR resonances with temperature (Figure~\ref{fig:Temperature}a), quantified as a decrease in the inhomogeneous dephasing time $T_2^*$ (Figure~S4). In a conventional scenario, $T_2^*$  is expected to remain nearly temperature independent as long as $T_1 \gg T_2^*$, and to decrease only at elevated temperatures when $T_1$ approaches the nanosecond-long $T_2^*$ \cite{mikhailov2018electron}. In contrast, in our measurements $T_2^*$ shows a pronounced temperature dependence, even though $T_1 \gg T_2^*$. One possible explanation is the temperature dependence of the inhomogeneity in the system. This inhomogeneity may result in a broad distribution of $T_1$ in the spin ensemble. At low temperature, the measured signal is dominated by the carriers with the longest $T_1$ [see Equation~\eqref{eq:SI}], which could correspond to a narrow distribution of precession frequencies and thus a longer $T_2^*$. Increase in the temperature primarily leads to the suppression of the longest $T_1$ and extension of the spin ensemble that dominates the ODMR signal. The larger ensemble shows a larger spread of the Larmor frequencies, resulting in shorter $T_2^*$. This can also explain the different activation energies for $T_1$ and $T_2^*$.

In both samples we also find multiple spin resonances which accompany the main electron and hole peaks. These resonances show up under specific conditions, namely at low fields and Larmor precession frequencies (Figure~\ref{fig:fine0.4}a) and at laser energies below the exciton resonance  (Figure~\ref{fig:fine0.4}c). Resonances that have $g$ factors close to that of electron (hole) are attributed to electron (hole) spin subensembles and labeled as $e_i$ ($h_i$). Electron satellite resonances show an offset in the dependence of their frequency on magnetic field (Figures~\ref{fig:fine0.4}b, S8b and Tables~\ref{tab:1},~\ref{tab:2}). One can attribute these offsets to internal fields, locally experienced by the corresponding carrier spin subensembles. In particular, the offset may be related to the effective field arising from the exchange interaction in an exciton \cite{belykh2022submillisecond} or to the Overhauser field of the nuclear spin fluctuations \cite{ meliakov2024hole,meliakov2026hyperfine}. However, these cases would be characterized by a positive offset, i.e., a finite frequency at $B=0$ which is the case only for the $e_1$ resonance in Figure~\ref{fig:fine0.4}b. One should mention also the dynamic nuclear polarization, where optical pumping with certain helicity results in the Overhauser field directed along the external field, but independent of its sign. An internal field $\mathbf{B}_\text{i}$, if it is independent of the sign of the external field ($\mathbf{B}$ vs $-\mathbf{B}$), should effectively shift the magnetic field dependence horizontally. However, this interpretation is contradictory to the experimental observations: the corresponding dependence is symmetric when the magnetic field is scanned from negative to positive values and is shifted vertically to lower frequencies (Figure S5b in the Supporting Information). This may take place if the internal magnetic field changes sign with the external magnetic field. 

We emphasize the ultralong spin relaxation time of these satellite resonances, which becomes further enhanced for decreasing energy of the corresponding optical transition (Figure~\ref{fig:fine0.4}c). The longest $T_1$ reaches 2.1~ms for the $h_1$ transition in the MA$_{0.8}$FA$_{0.2}$PbI$_3$ sample (see Figure S8d and Table~\ref{tab:2}). The fact that these resonances are observed at decreasing laser energy suggests their localization character. On the other hand, for strongly localized carriers the effect of random nuclear spins is expected to be enhanced. This should lead to a decreasing $T_1$, especially at low fields (Figure~\ref{fig:odmr}d) and to broadened ODMR resonances. In fact, all new resonances, although appearing at low $B$, are characterized by an exceptionally long $T_1$ and a narrow ODMR spectrum, even those having hole character (Figures~\ref{fig:fine0.4}c and S8a). We note that in the extreme limit of strong localization, a carrier may strongly interact with only a few nuclei. This interaction, instead of causing spin-resonance broadening, can lead to the formation of several discrete (and narrow) spin energy levels in the interacting carrier-nuclei system. This scenario is observed for strongly localized spin states, e.g., in rare earth ions \cite{abragam2012electron,lovric2011hyperfine}.

\begin{figure}
        \centering
        \includegraphics[width=0.9\linewidth]{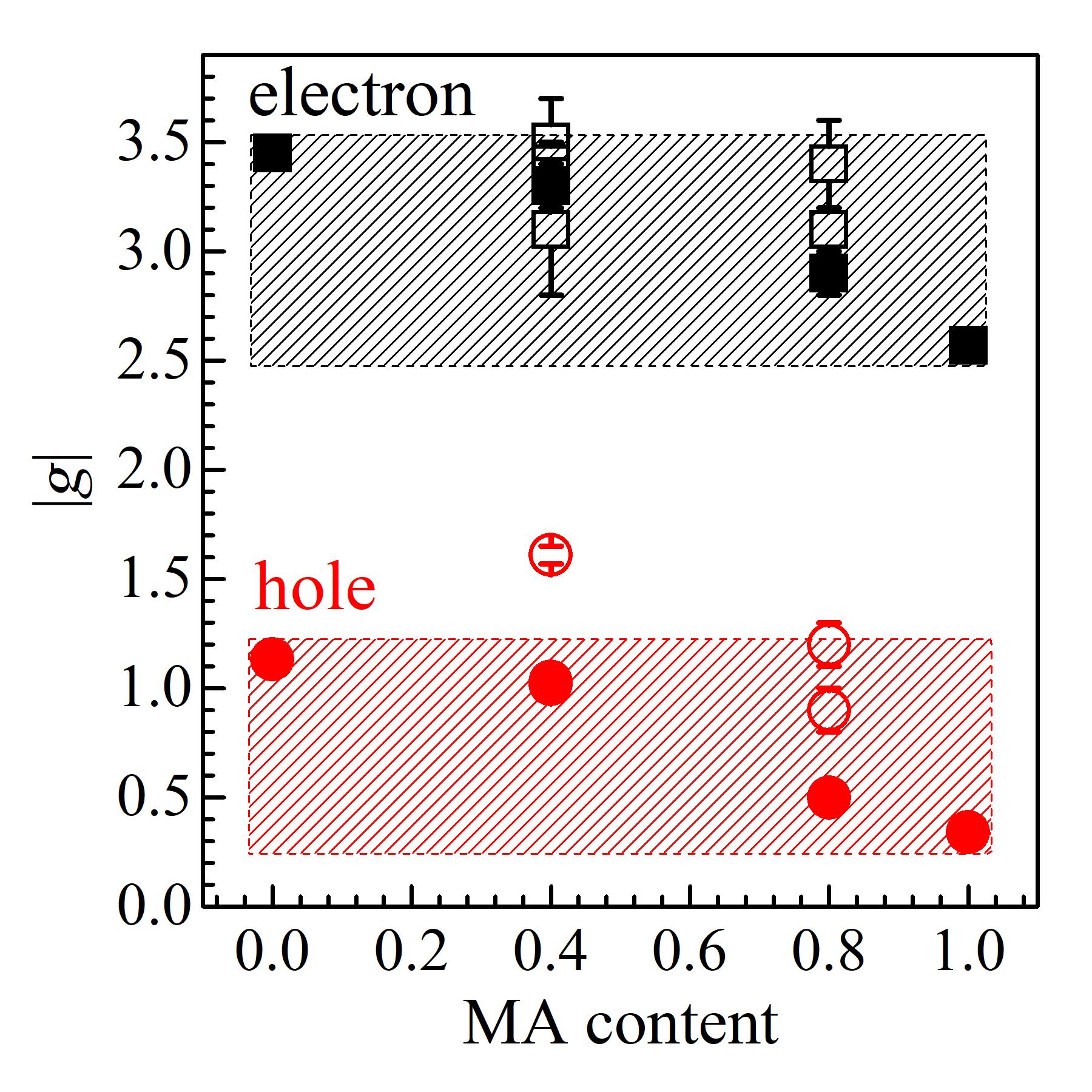}
        \caption{Absolute value of $g$ factors for electrons (black squares) and holes (red circles) as a function of the MA content in MA$_{x}$FA$_{1-x}$PbI$_3$ crystals. For FAPbI$_3$ crystal the data is taken from the work \cite{lague2024spin}. For MAPbI$_3$ crystal with anisotropic $g$ factor we take the lowest $|g|$ values from the work \cite{kirstein2022spin}. Open symbols show $|g|$ for the satellite resonances. Dashed regions show ranges of $g$ factors for MA content spanning from 0 to 1.}
        \label{fig:MA}
    \end{figure}

We note that in the FAPbI$_3$ the electron and hole $g$ factors  at the bandgap energy of 1.48~eV are $g_\text{e} = 3.45$ and $g_\text{h} = -1.13$  \cite{lague2024spin}. In the MAPbI$_3$ the electron and hole $g$ factors are anisotropic and span the range $g_\text{e} = 2.5\div3.0$, $g_\text{h} = -(0.3\div0.7)$ at the bandgap energy of 1.637~eV \cite{kirstein2022spin,kopteva2024weak}. It is instructive to compare these $g$ factors to those presented in the Tables~\ref{tab:1} and \ref{tab:2} for the MA$_x$FA$_{1-x}$PbI$_3$ compounds. For the quasi-free electron state e $|g_\text{e}| = 3.3$ for MA$_{0.4}$FA$_{0.6}$PbI$_3$, which is closer to the value in FAPbI$_3$ and $|g_\text{e}| = 2.9$ for MA$_{0.8}$FA$_{0.2}$PbI$_3$ which is closer to the value in MAPbI$_3$. The same tendency holds for the hole h state, for which $|g_\text{h}| = 1.0$ for MA$_{0.4}$FA$_{0.6}$PbI$_3$  and $|g_\text{h}| = 0.5$ for MA$_{0.8}$FA$_{0.2}$PbI$_3$. The values of $|g|$ as a function of MA content are summarized in Figure~\ref{fig:MA}. The $g$ factors for satellite electron resonances e$_1$, e$_2$, e$_3$ are close to that of e and within measurement error fall within the range from 2.5 to 3.45 corresponding to an MA content from 1 to 0. For the hole-like satellites h$_2$ has $g$ factor within the expected $g$ factor range of MA content from 1 to 0, while h$_1$ has a $|g|$ value larger than expected. The satellite states can be related to the crystal defects having fluctuations of MA content, different type or orientation of the crystal lattice. We note that carrier localization in such sites can lead to significant deviations of the $g$ factor from the value predicted by the bandgap energy \cite{nestoklon2023tailoring,meliakov2025lande}.

In fact, the relative number of these satellite states is quite small, and they are revealed in the resonance spin inertia technique due to their long $T_1$ time. Indeed, the signal in our experiment is proportional to $T_1^2$ [Equation~\eqref{eq:SI}]~\cite{belykh2022selective,belykh2026amr}. Considering that the ratio of the $T_1$ times between the satellite and the main resonances is about $10-100$, we can conclude that the corresponding ratio of spin populations is about $10^{-4} - 10^{-2}$. 
This complicates the observation of these satellite resonances by other techniques, e.g.,  time-resolved Faraday/Kerr rotation.

\section{Conclusions}

In summary, we have conducted a comprehensive ODMR investigation of the spin properties in mixed-A-site hybrid organic–inorganic perovskite MA$_x$FA$_{1-x}$PbI$_3$ single crystals with \textit{x} = 0.4 and 0.8. 
Across a wide range of magnetic fields and optical transition energies, we have observed electron and hole resonances with $g$ factor absolute values of 3.3 (2.9) and 1.0 (0.5) for electrons and holes, respectively, in the sample with $x = 0.4$ (0.8). The widths of the ODMR peaks allow one to evaluate Overhauser fields of approximately $0.5-0.8$~mT for electrons and $4-12$~mT for holes. These resonances have longitudinal spin relaxation times $T_1$ reaching hundreds of microseconds, which confirms the absence of the Dyakonov-Perel spin relaxation mechanism due to the lattice symmetry in this system. By analyzing the magnetic-field dependence of $T_1$, we have evaluated nuclear field correlation times $\tau_\text{c}$, which are about $0.04$ $\mu$s for electrons and $0.9$~$\mu$s for holes in MA$_{0.4}$FA$_{0.6}$PbI$_3$ crystal and about order of magnitude longer in MA$_{0.8}$FA$_{0.2}$PbI$_3$ crystal. These times are dominated by carrier hopping in a weak localizing potential landscape and indicate that carriers are more delocalized in the sample with reduced MA content. The carriers are also delocalized by increasing the temperature from 1.6 to 7 K, which leads to a moderate decrease in $T_1$ and pronounced decrease in the resident carriers density.

At low rf field frequencies (and magnetic fields), we have resolved a set of carrier spin subensembles, each with a distinct $g$-factor absolute values spanning the range of $2.9-3.5$ for electrons and of $0.5-1.6$ for holes. The relative amplitude of the different peaks strongly depends on the optical transition energy, suggesting their origin from different subensembles of electrons and holes with different degrees of localization and distinct hyperfine environments. Furthermore, all detected carrier subensembles exhibit micro‐to‐millisecond $T_1$ times, with a record value of 2.1~ms, underscoring the exceptionally slow spin relaxation in mixed-cation hybrid perovskite single crystals. These findings establish hybrid organic–inorganic perovskite single crystals as a compelling solid-state platform for spin physics, revealing a complex interplay of \textit{g}-factor dispersion, carrier localization, and hyperfine interaction, and combining long spin lifetimes with optical control in a technologically relevant material system. 


\section{Experimental Section}
\textbf{Samples}

The MA$_{0.4}$FA$_{0.6}$PbI$_3$ and MA$_{0.8}$FA$_{0.2}$PbI$_3$ single crystals were synthesized according to the well-established inverse crystallization  method~\cite{chen2019single,alsalloum2020low}.  This method reliably yields high-quality single crystals with well-defined facets and long carrier diffusion lengths~\cite{turedi2022single}. X-ray diffraction measurements further confirm the excellent structural quality of these perovskite single crystals~\cite{yang2022engineering}. The studied MA$_x$FA$_{1-x}$PbI$_3$ single crystals with $x = 0.4$ and 0.8 were synthesized from appropriately mixed MAI, FAI, and PbI$_2$ perovskite precursors. The precursors were injected between two polytetrafluoroethylene coated glasses and slowly heated to 120\textsuperscript{◦}C. The samples have square shapes that reach about 2×2 mm in the (001) crystallographic plane and a thickness of about 30~$\mu$m. 

\textbf{Optical measurements}

The photoluminescence of the crystals is dispersed by a 0.5 m monochromator and detected with a charge-coupled-device (CCD) camera following its excitation with a 3.06 eV continuous-wave (cw) diode laser. Reflectivity spectra were measured using a halogen lamp in back-reflection geometry. 

\textbf{ODMR measurements}

The experimental scheme used to measure the spin resonances and the longitudinal spin relaxation time $T_1$ is shown in Figure~\ref{fig:PL, abs and setup}b. We use a Coherent Chameleon Discovery laser system emitting 100~fs pulses at a repetition rate of 80~MHz, which wavelength is tunable over a wide spectral range. To reduce the spectral width to $\sim 1$~nm, a grating-based pulse shaper is used. The sample is placed in a He-bath cryostat with superconducting coils. At $T < 4.2$~K the sample was immersed in a liquid helium, while at higher temperatures it was held in a helium gas. The magnetic field \textit{B} is applied parallel to the sample normal ($\mathbf{B} \parallel \mathbf{k}$) (Faraday geometry). Optical spin orientation and spin polarization probing are performed using the same laser beam with elliptical polarization. The circular polarization component of the beam serves as pump for the carrier spins, while the linear component is used to probe the spin polarization via the Kerr rotation effect~\cite{belykh2021stimulated,belykh2022submillisecond,belykh2026amr}. The Kerr rotation is detected using a Wollaston prism, splitting the beam into two orthogonally polarized beams of approximately equal intensity which are detected by a balanced photodetector. The rf magnetic field is applied using a small coil near the sample surface. The current through the coil is driven by a function generator, which creates a sinusoidal voltage with frequency  $f_{\rm rf}$, ranging from 100 to 4500 MHz. The generator output is modulated sinusoidally at frequency $f_\text{mod}$, ranging from 0.1 to 100 kHz for synchronous detection with a lock-in amplifier. Thus, we measure the ODMR signal as a difference between Kerr rotation amplitude with the rf field at low and high level, which in turn is proportional to the corresponding difference $\Delta S$ in the spin polarization. The spin polarization created by resonant laser excitation accumulates along the external magnetic field $\textbf{B}$. The spins are addressed by the rf field only if their Larmor precession frequency $f_{\rm L}$ matches  $f_{\rm rf}$. To implement the resonant spin inertia method, the rf field modulation frequency $f_\text{mod}$ is varied. As  $f_\text{mod}$ increases, the corresponding modulation period becomes shorter than $T_1$, leading to a measurable reduction in the spin modulation amplitude. This dependence allows us to evaluate the $T_1$ time by means of Equation~\eqref{eq:SI} \cite{belykh2022selective}.

\medskip
\textbf{Supporting Information} 
Generalization of the spin inertia equation for two time scales, magnetic field dependencies of the spin dephasing time and the width of the ODMR peak in the MA\textsubscript{0.4}FA\textsubscript{0.6}PbI\textsubscript{3} crystal, dependence of the longitudinal spin relaxation time of electrons and holes on laser power, temperature dependencies of the spin dephasing times in the MA\textsubscript{0.4}FA\textsubscript{0.6}PbI\textsubscript{3} crystal, ODMR signals as a function of the magnetic field scanning from negative to positive values, spin dephasing time of holes as a function of laser excitation energy in MA\textsubscript{0.4}FA\textsubscript{0.6}PbI\textsubscript{3} crystal, spin relaxation times of different electron and hole subensembles as a functions of magnetic field, ODMR investigation of the MA$_{0.8}$FA$_{0.2}$PbI$_3$ single crystal.

\par 
Supporting Information is available from the Wiley Online Library or from the author.

\medskip
\textbf{Acknowledgements} \par 
We acknowledge the financial support by the Deutsche Forschungsgemeinschaft (project YA 65/28-1, no. 527080192).  The work at ETH Z\"urich (B.T. and M.V.K.) was financially supported by the Swiss National Science Foundation (grant agreement 200020E 217589, funded through the DFG-SNSF bilateral program), and by the ETH Z\"urich through the ETH+ Project SynMatLab.

\newpage
\onecolumngrid
\setcounter{page}{1}
\setcounter{section}{0}
\setcounter{figure}{0}
\setcounter{table}{0}
\setcounter{equation}{0}
\renewcommand{\thesection}{S\arabic{section}}
\renewcommand{\thesubsection}{S\arabic{section}.\arabic{subsection}}
\renewcommand{\thepage}{S\arabic{page}}
\renewcommand{\thefigure}{S\arabic{figure}}
\renewcommand{\thetable}{S\arabic{table}}
\renewcommand{\theequation}{S\arabic{equation}}

\renewcommand{\bibnumfmt}[1]{[S#1]}
\renewcommand{\citenumfont}[1]{S#1}

\textbf{\large Supporting Information: Millisecond spin relaxation times of distinct electron and hole subensembles in MA$_x$FA$_{1-x}$PbI$_3$ perovskite crystals}\\[6pt]

\textbf{S1. Generalization of the spin inertia equation for two carriers subensembles}  

For the case of two electron or hole subensembles having similar \textit{g}-factors, but different spin relaxation times $T_{1,1}$ and $T_{1,2}$, the equation describing the resonant spin inertia derived in Ref.~\cite{belykh2022selective}  can be generalized. We detect the spin signal in the $X$ and $Y$ channels of the lock in amplifier in phase with the rf field modulation and with a $\pi/2$ phase shift, respectively. These signals should add linearly for the two subensembles:
\begin{equation}
X=\sum_{i=1}^{2}\frac{A_{i}T_{1i}^2}{1+4\pi^2T_{1i}^2f_{\mathrm{m}}^2}
\tag{S1} 
\end{equation}
\begin{equation}
 Y=\sum_{i=1}^{2}\frac{2\pi T_{1i}^2f_{m}^2A_{i}}{1+4\pi^2T_{1i}^2f_{\mathrm{m}}^2}
 \tag{S2} 
\end{equation}

The total signal depicted in Figure~2c is calculated as
\begin{equation}
S=\sqrt{X^2+Y^2}.
\tag{S3} 
\end{equation}
Then, the average longitudinal spin relaxation time $T_1$ can be calculated via the intensity-weighted method, using the following equation 
\begin{equation}
 T_{1,avg}=\frac{\sum_{i=1}^{2}A_{i}T_{1i}^{2}}{\sum_{i=1}^{2}A_{i}T_{1i}}.
\tag{S4} 
\end{equation}
To visualize the contributions of the individual components, we plot the two components separately together with the resulting spin inertia curve in Figure S1.

\begin{figure}[H]
        \centering
        \includegraphics[width=0.75\linewidth]{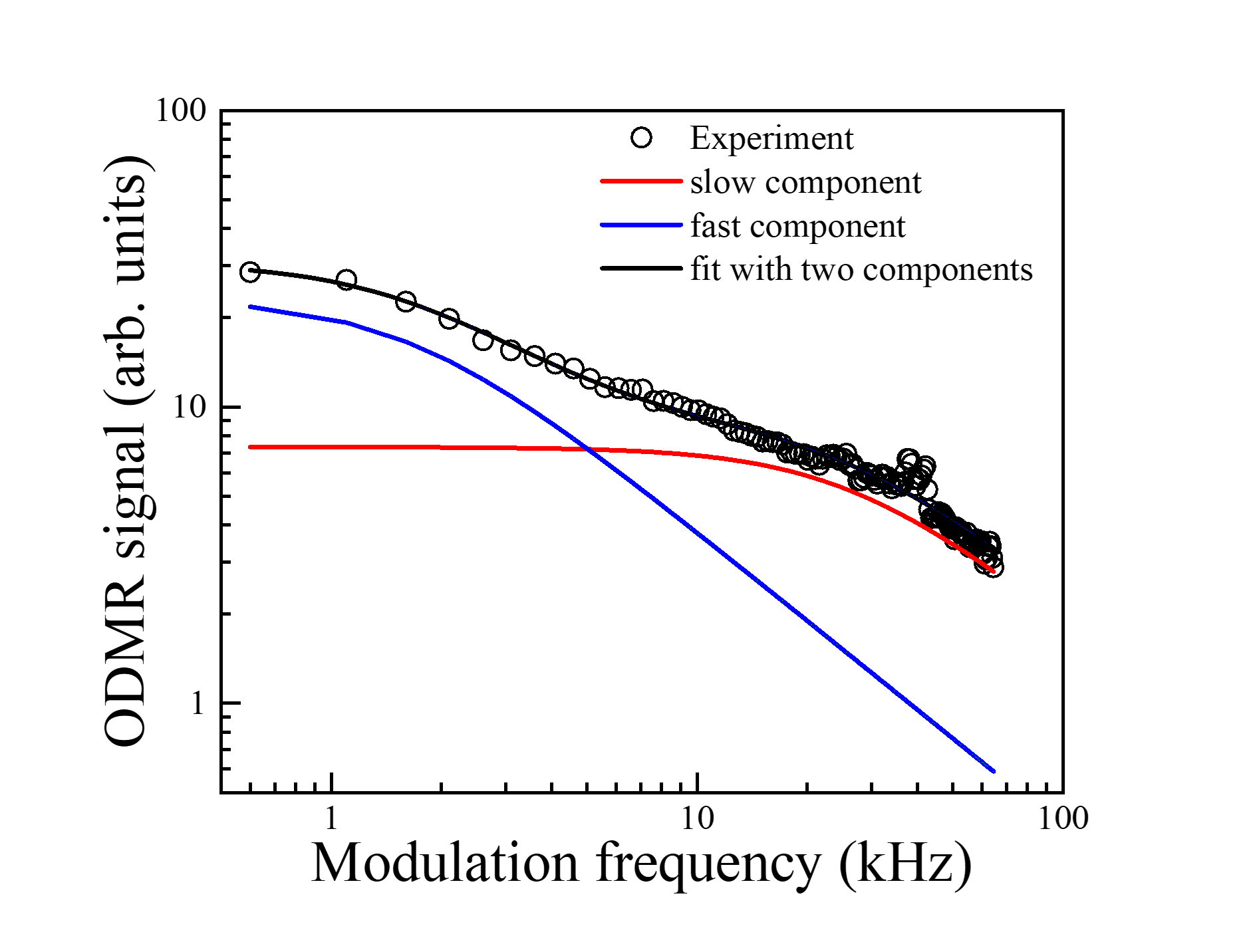}
        \caption{ODMR signal as function of the modulation frequency, $f_\text{mod}$, for holes in the MA\textsubscript{0.4}FA\textsubscript{0.6}PbI\textsubscript{3} crystal at $B = 106$~mT. The black line is a fit with two components using the equations (S1-S3). The blue and red lines show the individual contributions of the slow and fast components, respectively. $T = 1.6$~K.}
        \label{fig:fit}
    \end{figure}

\textbf{S2. Magnetic field dependencies of the spin dephasing time and the width of the ODMR peak in the MA\textsubscript{0.4}FA\textsubscript{0.6}PbI\textsubscript{3} crystal.}   

We can analyze the width of electron and hole in the ODMR resonances, $\Delta B$, defined as standard deviation of the Gaussian distribution. The width $\Delta B$ is contributed by the spread on nuclear fields $\Delta_\text{N}$ and by the spread of $g$ factors $\Delta g$ \cite{belykh2026millisecond} and can be described by the equation:
\begin{equation}
\Delta B = \left[\Delta_\text{N}^2/2 + \left(\frac{\Delta g}{g} B \right)^2\right]^{1/2}.
\tag{S5} 
\label{eq:dB}
\end{equation}
This equation assumes the Gaussian distributions of $g$ factors and nuclear fields.
Figure~S2 summarizes $\Delta B$, extracted from the ODMR spectra shown in Figure~2a of the main text at different rf frequencies $f_{\mathrm{rf}}$ in the MA\textsubscript{0.4}FA\textsubscript{0.6}PbI\textsubscript{3} crystal at $T = 1.6$~K. Figure~S2a shows the ODMR peak width $\Delta B$ of the electron and hole resonances as function of the magnetic field. The fact that $\Delta B$ weakly depends on $B$ indicates small contribution of the $g$ factor spread. Thus, ODMR peak width, in this range of $B$, is mainly determined by the nuclear field fluctuations.

Using Equation~(2) in the main text, the corresponding spin dephasing time $T_2^*$ can be evaluated. Figure~S2b presents the corresponding spin dephasing times $T_{2,\text{e}}^*$ and $T_{2,\text{h}}^*$ as function of the magnetic field. 

\begin{figure}[H]
        \centering
        \includegraphics[width=1\linewidth]{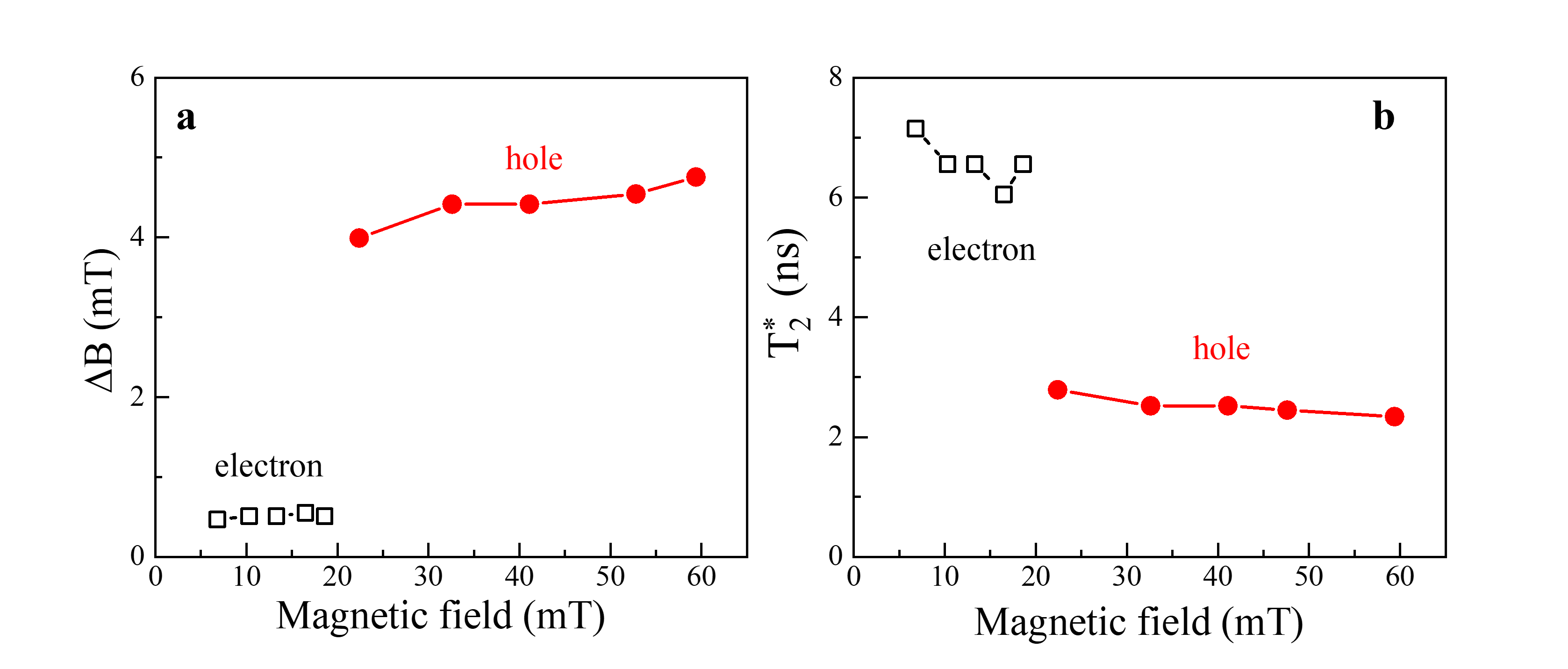}
        \caption{a) Width of the ODMR peak $\Delta B$ of the electron and hole resonances as function of the magnetic field in the MA\textsubscript{0.4}FA\textsubscript{0.6}PbI\textsubscript{3} crystal. b) Corresponding spin dephasing times of the electron and hole resonances as function of the magnetic field. The excitation laser energy is 1.528 eV. The laser power is 1 mW. $T = 1.6$~K. }
        \label{fig:T2 star}
    \end{figure}

\textbf{S3. Dependence of the longitudinal spin relaxation time of electrons and holes, \textit{T}\textsubscript{1}, on laser power.}

To obtain the spin relaxation time $T_1$ of the undisturbed spin system, we measured the ODMR signal as function of the modulation frequency for different laser powers at the fixed magnetic field strengths of $B = 31.8$ and $106$~mT. Equations~(S1)-(S3) provides a reasonable fit to the experimental dependencies, allowing us to extract $T_1$ for both electrons and holes at each laser power. The extracted $1/T_1$ values increase with increasing laser power, indicating that the optical excitation perturbs the spin system and accelerates the spin relaxation. By extrapolating the power dependence to the limit of zero laser power, we obtain the intrinsic spin relaxation times of the undisturbed system, $T_{1,\text{e}} = 22~\mu$s for electrons and $T_{1,\text{h}} = 88~\mu$s for holes.

\begin{figure}[H]
    \centering
    \includegraphics[width=1\linewidth]{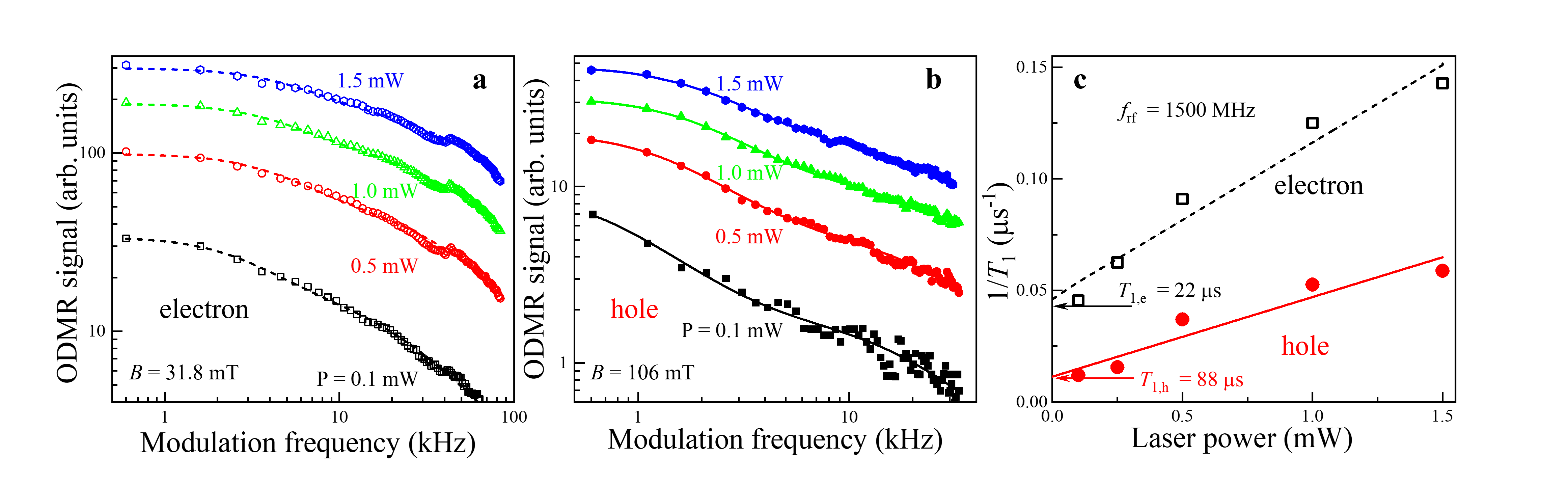}
    \caption{ODMR signal of a) electrons and b) holes in theMA\textsubscript{0.4}FA\textsubscript{0.6}PbI\textsubscript{3} crystal as function of the modulation frequency at $B = 31.8$~mT and 106~mT, respectively, for different laser powers. $f_{\rm rf} = 1500$~MHz. The lines show fits of the experimental data with Equations~(S1-S3). c) Laser power dependence of the longitudinal spin relaxation rate 1/\textit{T}\textsubscript{1}. The lines represent linear fits. The excitation energy is 1.528 eV. $T = 1.6$~K.  }
    \label{fig:S3}
\end{figure}

\textbf{S4. Temperature dependence of the spin dephasing time  $T_2^*$.}

With increasing temperature, the spin dephasing time $T_2^*$ decreases for both electrons and holes, as shown in Figure S4. Within a purely phenomenological description, they follow an Arrhenius-like function, Equation~5, in main text. From the fit, the activation energies for electrons and holes (\textit{E}\textsubscript{A,e} = 3.7 meV and \textit{E}\textsubscript{A,h} = 1.3 meV) can be obtained.
\begin{figure}[H]
        \centering
        \includegraphics[width=0.75\linewidth]{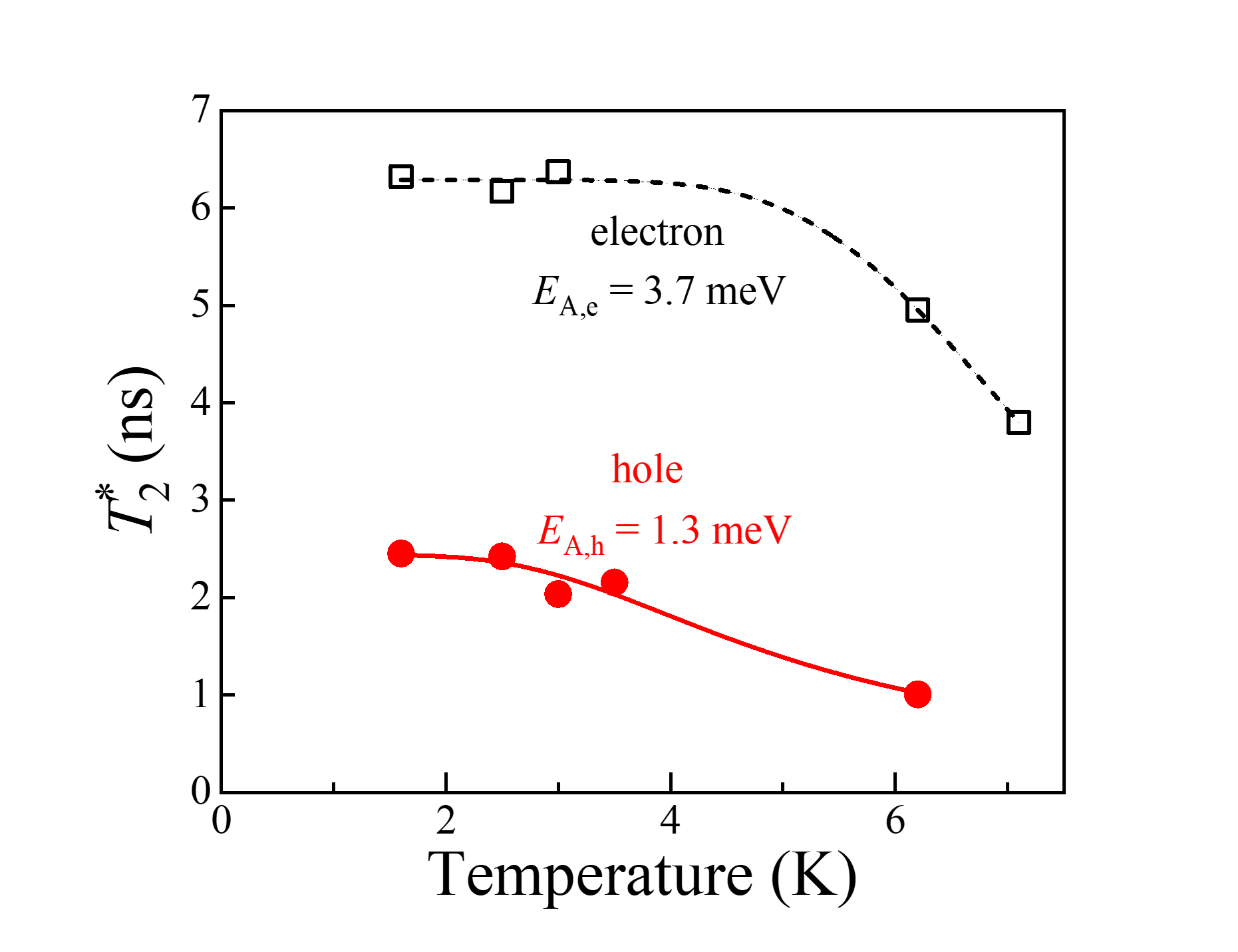}
        \caption{Temperature dependence of the spin dephasing time,  $T_2^*$, in the MA\textsubscript{0.4}FA\textsubscript{0.6}PbI\textsubscript{3} crystal. The lines are fits with Equation~(5) in the main text. The excitation energy is 1.528 eV. }
        \label{fig:s7}
    \end{figure}

\textbf{S5. ODMR signals as function of magnetic field, scanned from negative to positive direction.}

Figure~S5a shows the ODMR signal as function of magnetic field measured at different rf frequency on the MA\textsubscript{0.4}FA\textsubscript{0.6}PbI\textsubscript{3} crystal at $T = 1.6$~K, using the excitation laser energy of 1.524 eV. Several resonances corresponding to different carrier subensembles are clearly resolved. The magnetic field dependence of the corresponding resonance frequencies is presented in Figure~S5b, the lines show linear fits to the experimental data. Notably, the several resonances exhibit finite offsets in their frequency dependence on magnetic field. Such offsets are unlikely to originate from dynamical nuclear polarization \cite{kirstein2022lead}, because the dependencies remain symmetric when the magnetic field direction is reversed from negative to positive values, as demonstrated in Figure~S5b.

\begin{figure}[H]
    \centering
    \includegraphics[width=1\linewidth]{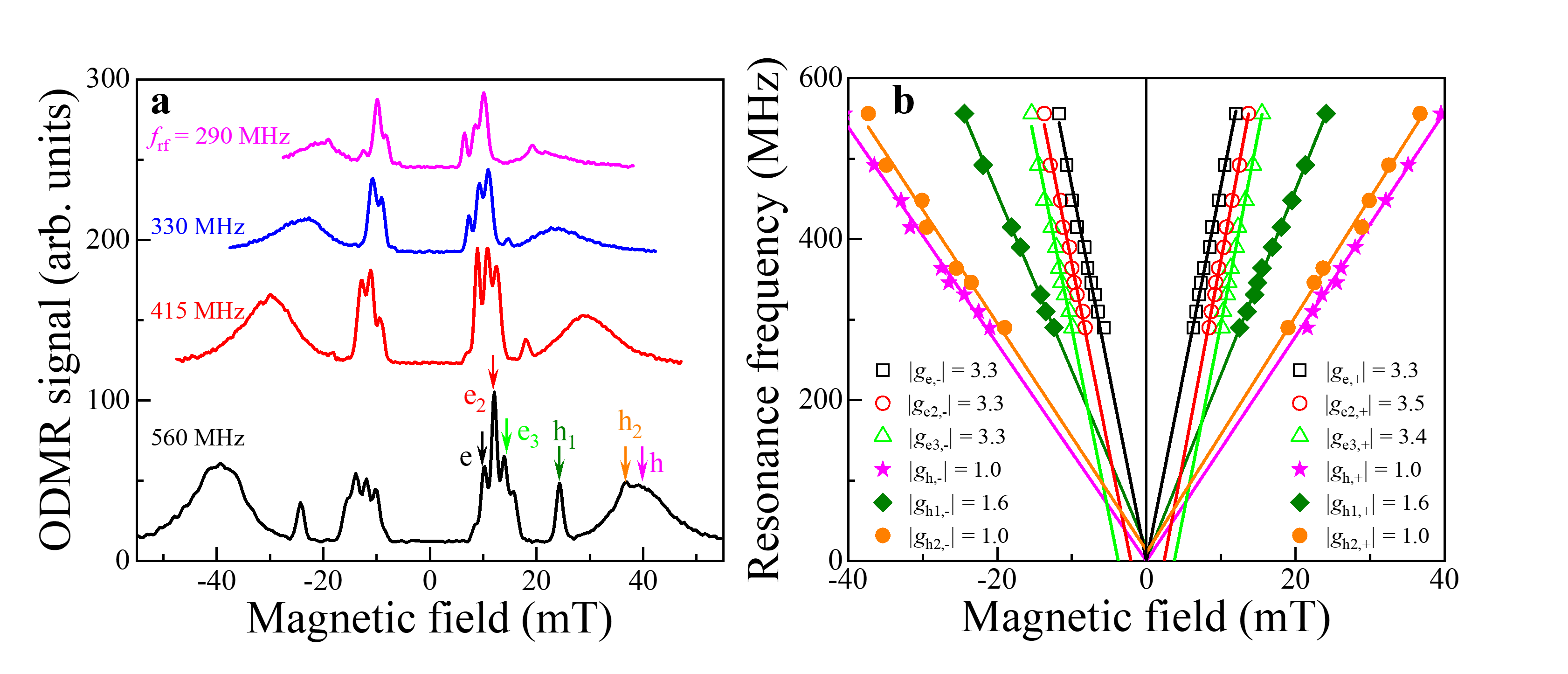}
    \caption{a) ODMR spectra of MA\textsubscript{0.4}FA\textsubscript{0.6}PbI\textsubscript{3} crystal measured at different rf frequencies. The curves are vertically shifted for clarity.  b) Magnetic-field dependence of the resonance frequencies corresponding to electrons and holes in the ODMR spectra, with linear fits. The excitation laser energy is 1.524 eV. The laser power is 1 mW. $T = 1.6$~K.} 
    \label{fig:S4}
\end{figure}

\textbf{S6. Spin dephasing time of holes as function of excitation energy.}

Figure~S6 shows the spin dephasing times $T_2^*$ of the h, h$_1$, and h$_2$ resonances in the MA\textsubscript{0.4}FA\textsubscript{0.6}PbI\textsubscript{3} crystal as a function of the excitation laser energy. The spin dephasing time estimated from the ODMR peak width reaches $T_2^* = 17$~ns for the h$_2$ spin subensemble at an excitation energy of 1.522~eV.

\begin{figure}[H]
        \centering
        \includegraphics[width=0.75\linewidth]{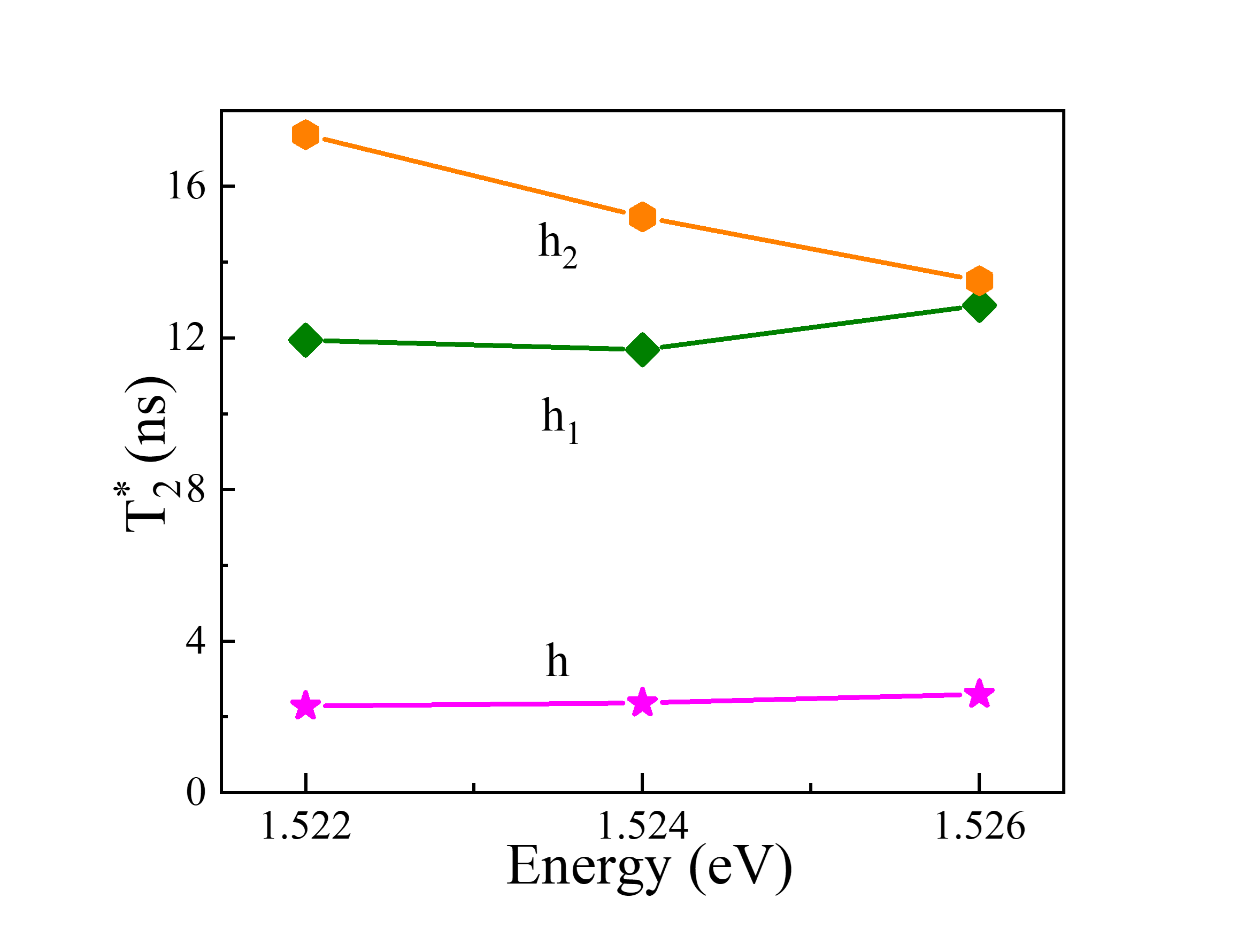}
        \caption{Spin dephasing times of the h h\textsubscript{1}, and h\textsubscript{2} resonances in the MA\textsubscript{0.4}FA\textsubscript{0.6}PbI\textsubscript{3} crystal as function of excitation laser energy. $f_{\rm rf} = 330$~MHz. The laser power is 2 mW. $T = 1.6$~K. The magnetic fields for the h\textsubscript{1}, h\textsubscript{2} and h resonances are 14.1, 21.3, and 24 mT, respectively. The lines are guides to the eye.}
        \label{fig:s6}
    \end{figure}
\textbf{S7. Spin relaxation times of different electron and hole subensembles as functions of  magnetic field.}

Figure S7a shows ODMR signal as a function of the modulation frequency at different magnetic fields, corresponding to the electrons and holes resonances at fixed rf frequency of 560~MHz.  From the fits, the spin relaxation times for electrons and holes can be obtained. The resulting spin relaxation times for electrons and holes are plotted as a function of magnetic field in Figure S7b.

\begin{figure}[H]
    \centering
    \includegraphics[width=1\linewidth]{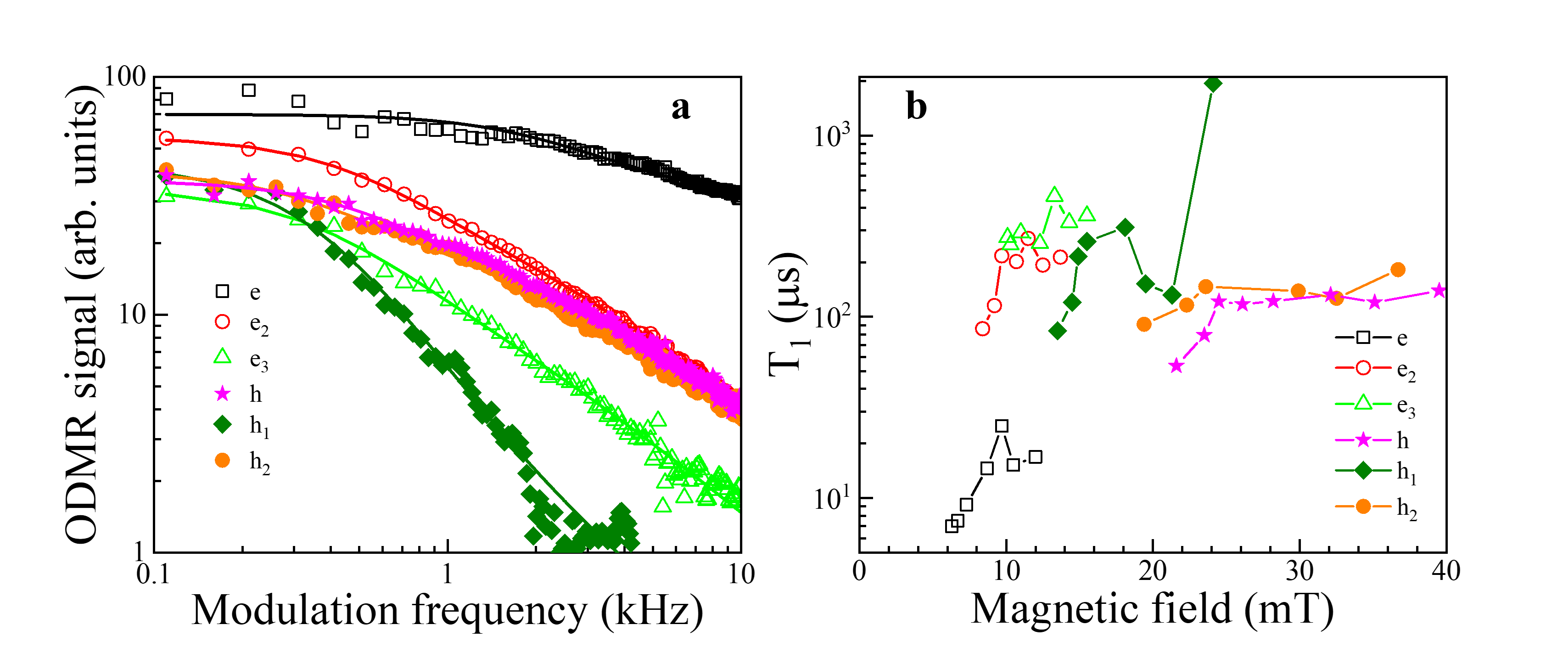}
    \caption{a) ODMR signals of MA\textsubscript{0.4}FA\textsubscript{0.6}PbI\textsubscript{3} crystal as a function of the modulation frequency at different magnetic fields, corresponding to the electron and hole resonances, at the fixed rf frequency of 560~MHz. The lines show fits of the experimental data with Equations (S1)-(S4). b) Spin relaxation times for different electron and hole subensembles as a function of the magnetic field. The excitation laser energy is 1.524 eV. The laser power is 1~mW. $T = 1.6$~K.}. 
    \label{fig:S4}
\end{figure}
    
\textbf{S8. ODMR investigation in MA$_{0.8}$FA$_{0.2}$PbI$_3$ single crystal.}
\begin{figure}[H]
        \centering
        \includegraphics[width=0.9\linewidth]{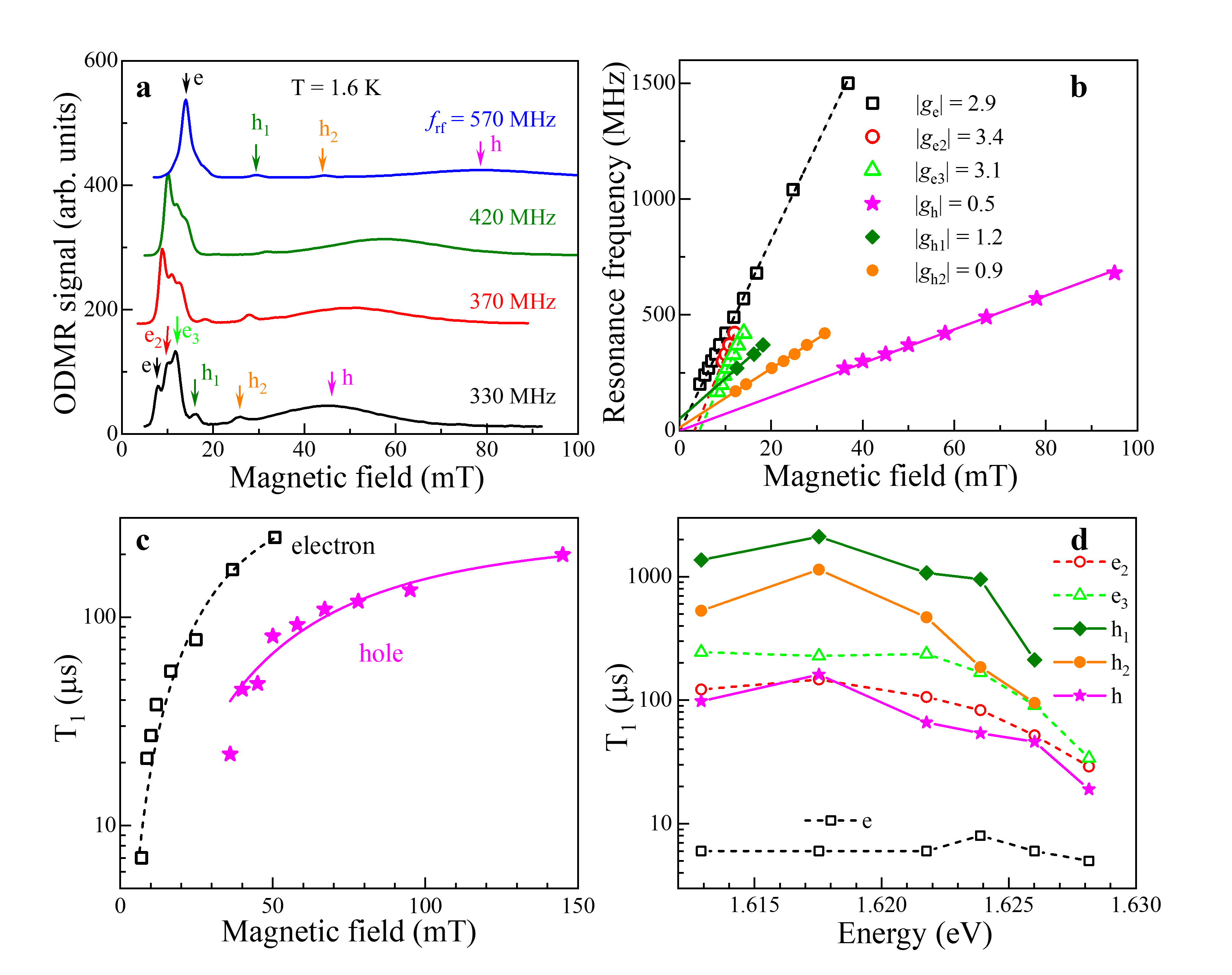}
        \caption{a) ODMR spectra of the MA$_{0.8}$FA$_{0.2}$PbI$_3$ crystal measured at different rf frequencies. The curves are vertically shifted for clarity. b) Magnetic-field dependence of the resonance frequencies corresponding to the electron and hole positions in the ODMR spectra, with the associated linear fits shown by the solid lines. c) Spin relaxation times $T_1$, for electrons and holes as  function of magnetic field. The lines show fits of the experimental data using Equation~(4) in the main text.  d)  $T_1$ as function of the laser photon energy for different subensembles of electrons and holes with the fixed rf frequency of 330~MHz. The laser power is 2~mW,  $T = 1.6$~K. The laser energy in (a),(b),(c) is 1.626~eV.}
        \label{fig:s7}
    \end{figure}
    
Multiple carrier resonances are also found for the MA$_{0.8}$FA$_{0.2}$PbI$_3$ crystal as shown in Figure~S8. With decreasing rf frequency, the number of peaks in the ODMR spectra increases. We determine the $g$-factors, corresponding to the different ODMR peaks from the slope of the linear fit of the resonance frequency dependence on the magnetic field shown in Figure~S8b. We assign the observed spin components to distinct carrier subensembles: the electron subensembles e, e\textsubscript{2}, and e\textsubscript{3} with $g$-factors of 2.9, 3.4, and 3.1, respectively, and the hole subensembles h, h\textsubscript{1}, and h\textsubscript{2} with $g$-factors of 0.5, 1.2, and 0.9, respectively. The energy dependence of $T_1$ for different subensembles in this sample is shown in Figure~S8d. A decrease of the laser photon energy leads to an increase of the spin relaxation times for all carrier spin subensembles, similar to the tendency found in the MA$_{0.4}$FA$_{0.6}$PbI$_3$ crystal (Figure~4d in the main text). For the h\textsubscript{1} hole spin subensemble, $T_1$ can reach 2 ms at the laser energy of 1.617~eV. The other carrier subensembles exhibit faster relaxation, reflecting varying degrees of localization and different Overhauser fields.


\begin{thebibliography}{10}
\providecommand{\url}[1]{\texttt{#1}}
\providecommand{\urlprefix}{URL }

\bibitem{herz2017charge}
L. M. Herz,
``Charge-carrier mobilities in metal halide perovskites: Fundamental mechanisms and limits'',
\href{https://doi.org/10.1021/acsenergylett.7b00276}{ACS Energy Lett. \textbf{2}, 1539--1548 (2017)}.

\bibitem{Jeong2021}
J. Jeong, M. Kim, J. Seo, H. Lu, P. Ahlawat, A. Mishra, Y. Yang, M. A. Hope, F. T. Eickemeyer, M. Kim, Y. J. Yoon, I. W. Choi, B. P. Darwich, S. J. Choi, Y. Jo, J. H. Lee, B. Walker, S. M. Zakeeruddin, L. Emsley, U. Rothlisberger, A. Hagfeldt, D. S. Kim, M. Gr{\"a}tzel, and J. Y. Kim,
``Pseudo-halide anion engineering for $\alpha$-FAPbI$_3$ perovskite solar cells'',
\href{https://doi.org/10.1038/s41586-021-03406-5}{Nature \textbf{592}, 381--385 (2021)}.

\bibitem{he2023single}
J. He, D. Li, H. Liu, J. Xiang, J. Bai, Y. Ren, Z. Wang, M. Xia, X. Yin, L. Yuan, F. Zhang, and S. Wang,
``Single-crystal seeds inducing the crystallization of high-performance $\alpha$-FAPbI$_3$ for efficient perovskite solar cells'',
\href{https://doi.org/10.1002/aenm.202300451}{Adv. Energy Mater. \textbf{13}, 2300451 (2023)}.

\bibitem{wei2019halide}
H. Wei, and J. Huang,
``Halide lead perovskites for ionizing radiation detection'',
\href{https://doi.org/10.1038/s41467-019-08981-w}{Nat. Commun. \textbf{10}, 1066 (2019)}.

\bibitem{nestoklon2018optical}
M. O. Nestoklon, S. V. Goupalov, R. I. Dzhioev, O. S. Ken, V. L. Korenev, Y. G. Kusrayev, V. F. Sapega, C. de Weerd, L. Gomez, T. Gregorkiewicz, J. Lin, K. Suenaga, Y. Fujiwara, L. B. Matyushkin, and I. N. Yassievich,
``Optical orientation and alignment of excitons in ensembles of inorganic perovskite nanocrystals'',
\href{https://doi.org/10.1103/PhysRevB.97.235304}{Phys. Rev. B \textbf{97}, 235304 (2018)}.

\bibitem{kopteva2024highly}
N. E. Kopteva, D. R. Yakovlev, E. Yalcin, I. A. Akimov, M. O. Nestoklon, M. M. Glazov, M. Kotur, D. Kudlacik, E. A. Zhukov, E. Kirstein, O. Hordiichuk, D. N. Dirin, M. V. Kovalenko, and M. Bayer,
``Highly-polarized emission provided by giant optical orientation of exciton spins in lead halide perovskite crystals'',
\href{https://doi.org/10.1002/advs.202403691}{Adv. Sci. \textbf{11}, 2403691 (2024)}.

\bibitem{becker2018bright}
M. A. Becker, R. Vaxenburg, G. Nedelcu, P. C. Sercel, A. Shabaev, M. J. Mehl, J. G. Michopoulos, S. G. Lambrakos, N. Bernstein, J. L. Lyons, T. St{\"o}ferle, R. F. Mahrt, M. V. Kovalenko, D. J. Norris, G. Rain{\`o}, and A. L. Efros,
``Bright triplet excitons in caesium lead halide perovskites'',
\href{https://doi.org/10.1038/nature25147}{Nature \textbf{553}, 189--193 (2018)}.

\bibitem{kirstein2022lead}
E. Kirstein, D. R. Yakovlev, M. M. Glazov, E. Evers, E. A. Zhukov, V. V. Belykh, N. E. Kopteva, D. Kudlacik, O. Nazarenko, D. N. Dirin, M. V. Kovalenko, and M. Bayer,
``Lead-dominated hyperfine interaction impacting the carrier spin dynamics in halide perovskites'',
\href{https://doi.org/10.1002/adma.202105263}{Adv. Mater. \textbf{34}, 2105263 (2022)}.

\bibitem{kirstein2025resonant}
E. Kirstein, D. R. Yakovlev, E. A. Zhukov, N. E. Kopteva, B. Turedi, M. V. Kovalenko, and M. Bayer,
``Resonant spin amplification and accumulation in MAPbI$_3$ single crystals'',
\href{https://doi.org/10.1002/advs.202502735}{Adv. Sci. \textbf{12}, 2502735 (2025)}.

\bibitem{kudlacik2024optical}
D. Kudlacik, N. E. Kopteva, M. Kotur, K. V. Kavokin, C. Harkort, M. Karzel, D. R. Yakovlev, E. A. Zhukov, E. Evers, V. V. Belykh, and M. Bayer,
``Optical spin orientation of localized electrons and holes interacting with nuclei in a FA$_{0.9}$Cs$_{0.1}$PbI$_{2.8}$Br$_{0.2}$ perovskite crystal'',
\href{https://doi.org/10.1021/acsphotonics.4c00637}{ACS Photonics \textbf{11}, 2757--2769 (2024)}.

\bibitem{belykh2019coherent}
V. V. Belykh, D. R. Yakovlev, M. M. Glazov, P. S. Grigoryev, M. Hussain, J. Rautert, D. N. Dirin, M. V. Kovalenko, and M. Bayer,
``Coherent spin dynamics of electrons and holes in CsPbBr$_3$ perovskite crystals'',
\href{https://doi.org/10.1038/s41467-019-08625-z}{Nat. Commun. \textbf{10}, 673 (2019)}.

\bibitem{kirstein2024coherent}
E. Kirstein, E. A. Zhukov, D. R. Yakovlev, N. E. Kopteva, E. Yalcin, I. A. Akimov, O. Hordiichuk, D. N. Dirin, M. V. Kovalenko, and M. Bayer,
``Coherent carrier spin dynamics in FAPbBr$_3$ perovskite crystals'',
\href{https://doi.org/10.1021/acs.jpclett.4c00098}{J. Phys. Chem. Lett. \textbf{15}, 2893--2903 (2024)}.

\bibitem{siyushev2014coherent}
P. Siyushev, K. Xia, R. Reuter, M. Jamali, N. Zhao, N. Yang, C. Duan, N. Kukharchyk, A. D. Wieck, R. Kolesov, et al.,
``Coherent properties of single rare-earth spin qubits'',
\href{https://doi.org/10.1038/ncomms4895}{Nat. Commun. \textbf{5}, 3895 (2014)}.

\bibitem{belykh2021stimulated}
V. V. Belykh, A. R. Korotneva, and D. R. Yakovlev,
``Stimulated resonant spin amplification reveals millisecond electron spin coherence time of rare-earth ions in solids'',
\href{https://doi.org/10.1103/PhysRevLett.127.157401}{Phys. Rev. Lett. \textbf{127}, 157401 (2021)}.

\bibitem{w11v-2v4g}
M. Kotur, P. S. Bazhin, K. V. Kavokin, N. E. Kopteva, D. R. Yakovlev, D. Kudlacik, and M. Bayer,
``Dynamic polarization of nuclear spins by optically oriented electrons and holes in lead halide perovskite semiconductors'',
\href{https://doi.org/10.1103/PhysRevB.113.085204}{Phys. Rev. B \textbf{113}, 085204 (2026)}.

\bibitem{stano2022review}
P. Stano, and D. Loss,
``Review of performance metrics of spin qubits in gated semiconducting nanostructures'',
\href{https://doi.org/10.1038/s42254-022-00484-w}{Nat. Rev. Phys. \textbf{4}, 672--688 (2022)}.

\bibitem{belykh2022submillisecond}
V. V. Belykh, M. L. Skorikov, E. V. Kulebyakina, E. V. Kolobkova, M. S. Kuznetsova, M. M. Glazov, and D. R. Yakovlev,
``Submillisecond spin relaxation in CsPb(Cl,Br)$_3$ perovskite nanocrystals in a glass matrix'',
\href{https://doi.org/10.1021/acs.nanolett.2c01673}{Nano Lett. \textbf{22}, 4583--4588 (2022)}.

\bibitem{barak2022uncovering}
Y. Barak, I. Meir, J. Dehnel, F. Horani, D. R. Gamelin, A. Shapiro, and E. Lifshitz,
``Uncovering the influence of Ni$^{2+}$ doping in lead-halide perovskite nanocrystals using optically detected magnetic resonance spectroscopy'',
\href{https://doi.org/10.1021/acs.chemmater.1c03822}{Chem. Mater. \textbf{34}, 1686--1698 (2022)}.

\bibitem{dyakonov1972spin}
M.~I. Dyakonov, V.~I. Perel, ``Spin relaxation of conduction electrons in noncentrosymmetric semiconductors'', \newblock \emph{Soviet Physics–Solid State} \textbf{13} (1972): 3023--3026.


\bibitem{kopteva2025effect}
N. E. Kopteva, D. R. Yakovlev, E. Yalcin, I. V. Kalitukha, I. A. Akimov, M. O. Nestoklon, B. Turedi, O. Hordiichuk, D. N. Dirin, M. V. Kovalenko, and M. Bayer,
``Effect of crystal symmetry of lead halide perovskites on the optical orientation of excitons'',
\href{https://doi.org/10.1002/advs.202416782}{Adv. Sci. \textbf{12}, 2416782 (2025)}.

\bibitem{gribakin2026spin}
B. F. Gribakin, N. E. Kopteva, D. R. Yakovlev, I. A. Akimov, I. V. Kalitukha, B. Turedi, M. V. Kovalenko, and M. Bayer,
``Spin dynamics of excitons and carriers in mixed-cation MA$_x$FA$_{1-x}$PbI$_3$ perovskite crystals: Alloy fluctuations probed by optical orientation'',
\href{https://doi.org/10.1002/apxr.202500241}{Adv. Phys. Res., e00241 (2026)}.

\bibitem{pramanik2007observation}
S. Pramanik, C.-G. Stefanita, S. Patibandla, S. Bandyopadhyay, K. Garre, N. Harth, and M. Cahay,
``Observation of extremely long spin relaxation times in an organic nanowire spin valve'',
\href{https://doi.org/10.1038/nnano.2007.64}{Nat. Nanotechnol. \textbf{2}, 216--219 (2007)}.

\bibitem{mena2024room}
A. Mena, S. K. Mann, A. Cowley-Semple, E. Bryan, S. Heutz, D. R. McCamey, M. Attwood, and S. L. Bayliss,
``Room-temperature optically detected coherent control of molecular spins'',
\href{https://doi.org/10.1103/PhysRevLett.133.120801}{Phys. Rev. Lett. \textbf{133}, 120801 (2024)}.

\bibitem{kirstein2022spin}
E. Kirstein, D. R. Yakovlev, E. A. Zhukov, J. H{\"o}cker, V. Dyakonov, and M. Bayer,
``Spin dynamics of electrons and holes interacting with nuclei in MAPbI$_3$ perovskite single crystals'',
\href{https://doi.org/10.1021/acsphotonics.2c00096}{ACS Photonics \textbf{9}, 1375--1384 (2022)}.

\bibitem{heisterkamp2015longitudinal}
F. Heisterkamp, E. A. Zhukov, A. Greilich, D. R. Yakovlev,
V. L. Korenev, A. Pawlis, and M. Bayer,
\href{https://doi.org/10.1103/PhysRevB.91.235432}
{Phys. Rev. B \textbf{91}, 235432 (2015)}.

\bibitem{belykh2022selective}
V. V. Belykh, and S. R. Melyakov,
``Selective measurement of the longitudinal electron spin relaxation time T$_1$ of Ce$^{3+}$ ions in a YAG lattice: Resonant spin inertia'',
\href{https://doi.org/10.1103/PhysRevB.105.205129}{Phys. Rev. B \textbf{105}, 205129 (2022)}.

\bibitem{belykh2026amr}
V. V. Belykh,
``ODMR-Based Resonant Spin Inertia: A Sensitive Technique for Studying Spin Relaxation'',
\href{https://doi.org/10.1007/s00723-026-01858-4}{Appl. Magn. Reson. (2026)}.

\bibitem{chen2019single}
Z. Chen, B. Turedi, A. Y. Alsalloum, C. Yang, X. Zheng, I. Gereige, A. AlSaggaf, O. F. Mohammed, and O. M. Bakr,
``Single-crystal MAPbI$_3$ perovskite solar cells exceeding 21\% power conversion efficiency'',
\href{https://doi.org/10.1021/acsenergylett.9b00847}{ACS Energy Lett. \textbf{4}, 1258--1259 (2019)}.

\bibitem{alsalloum2020low}
A. Y. Alsalloum, B. Turedi, X. Zheng, S. Mitra, A. A. Zhumekenov, K. J. Lee, P. Maity, I. Gereige, A. AlSaggaf, I. S. Roqan, O. F. Mohammed, and O. M. Bakr,
``Low-temperature crystallization enables 21.9\% efficient single-crystal MAPbI$_3$ inverted perovskite solar cells'',
\href{https://doi.org/10.1021/acsenergylett.9b02787}{ACS Energy Lett. \textbf{5}, 657--662 (2020)}.

\bibitem{Yu16}
Z. G. Yu,
``Effective-mass model and magneto-optical properties in hybrid perovskites'',
\href{https://doi.org/10.1038/srep28576}{Sci. Rep. \textbf{6}, 28576 (2016)}.

\bibitem{Nestoklon21}
M. O. Nestoklon,
``Tight-binding description of inorganic lead halide perovskites in cubic phase'',
\href{https://doi.org/10.1016/j.commatsci.2021.110535}{Comput. Mater. Sci. \textbf{196}, 110535 (2021)}.

\bibitem{kirstein2022lande}
E. Kirstein, D. R. Yakovlev, M. M. Glazov, E. A. Zhukov, D. Kudlacik, I. V. Kalitukha, V. F. Sapega, G. S. Dimitriev, M. A. Semina, M. O. Nestoklon, M. V. Kovalenko, A. Baumann, J. H{\"o}cker, V. Dyakonov, and M. Bayer,
``The Land{\'e} factors of electrons and holes in lead halide perovskites: Universal dependence on the band gap'',
\href{https://doi.org/10.1038/s41467-022-30701-0}{Nat. Commun. \textbf{13}, 3062 (2022)}.

\bibitem{belykh2016large}
V. V. Belykh, D. R. Yakovlev, J. J. Schindler, E. A. Zhukov, M. A. Semina, M. Yacob, J. P. Reithmaier, M. Benyoucef, and M. Bayer,
``Large anisotropy of electron and hole g factors in infrared-emitting InAs/InAlGaAs self-assembled quantum dots'',
\href{https://doi.org/10.1103/PhysRevB.93.125302}{Phys. Rev. B \textbf{93}, 125302 (2016)}.

\bibitem{mikhailov2018electron}
A. V. Mikhailov, V. V. Belykh, D. R. Yakovlev, P. S. Grigoryev, J. P. Reithmaier, M. Benyoucef, and M. Bayer,
``Electron and hole spin relaxation in InP-based self-assembled quantum dots emitting at telecom wavelengths'',
\href{https://doi.org/10.1103/PhysRevB.98.205306}{Phys. Rev. B \textbf{98}, 205306 (2018)}.

\bibitem{Kirstein23}
E. Kirstein, D. S. Smirnov, E. A. Zhukov, D. R. Yakovlev, N. E. Kopteva, D. N. Dirin, O. Hordiichuk, M. V. Kovalenko, and M. Bayer,
``The squeezed dark nuclear spin state in lead halide perovskites'',
\href{https://doi.org/10.1038/s41467-023-42265-8}{Nat. Commun. \textbf{14}, 6683 (2023)}.

\bibitem{smirnov2018theory}
D. S. Smirnov, E. A. Zhukov, E. Kirstein, D. R. Yakovlev, D. Reuter, A. D. Wieck, M. Bayer, A. Greilich, and M. M. Glazov,
``Theory of spin inertia in singly charged quantum dots'',
\href{https://doi.org/10.1103/PhysRevB.98.125306}{Phys. Rev. B \textbf{98}, 125306 (2018)}.

\bibitem{kotur2026hyperfineinteractionelectronsholes}
M. Kotur, N. E. Kopteva, D. R. Yakovlev, B. Turedi, M. V. Kovalenko, and M. Bayer,
``Hyperfine interaction of electrons and holes with nuclei probed by optical orientation in MAPbI$_3$ perovskite crystals'',
\href{https://doi.org/10.1103/3ptm-4tpd}{Phys. Rev. B \textbf{113}, 245203 (2026)}.

\bibitem{Greilich2006}
A. Greilich, R. Oulton, E. A. Zhukov, I. A. Yugova, D. R. Yakovlev, M. Bayer, A. Shabaev, A. L. Efros, I. A. Merkulov, V. Stavarache, D. Reuter, and A. Wieck,
``Optical control of spin coherence in singly charged (In,Ga)As/GaAs quantum dots'',
\href{https://doi.org/10.1103/PhysRevLett.96.227401}{Phys. Rev. Lett. \textbf{96}, 227401 (2006)}.

\bibitem{Meier1984}
F.~Meier, B.~P. Zakharchenya, editors, \newblock \emph{Optical Orientation}, \newblock North-Holland, Amsterdam, (1984).


\bibitem{belykh2026millisecond}
V. V. Belykh, M. M. Glazov, S. R. Meliakov, D. R. Yakovlev, E. V. Kulebyakina, M. L. Skorikov, M. V. Kochiev, M. S. Kuznetsova, E. V. Kolobkova, and M. Bayer,
``Millisecond-long electron spin lifetime in CsPbI$_3$ perovskite nanocrystals revealed by optically detected magnetic resonance'',
\href{https://doi.org/10.48550/arXiv.2603.03038}{arXiv, 2603.03038 (2026)}.

\bibitem{meliakov2024hole}
S. R. Meliakov, V. V. Belykh, E. A. Zhukov, E. V. Kolobkova, M. S. Kuznetsova, M. Bayer, and D. R. Yakovlev,
``Hole spin precession and dephasing induced by nuclear hyperfine fields in CsPbBr$_3$ and CsPb(Cl,Br)$_3$ nanocrystals in a glass matrix'',
\href{https://doi.org/10.1103/PhysRevB.110.235301}{Phys. Rev. B \textbf{110}, 235301 (2024)}.

\bibitem{meliakov2026hyperfine}
S. R. Meliakov, E. A. Zhukov, V. V. Belykh, K. V. Kavokin, M. O. Nestoklon, E. V. Kulebyakina, M. L. Skorikov, E. V. Kolobkova, M. S. Kuznetsova, M. Bayer, and D. R. Yakovlev,
``Hyperfine interaction of electrons confined in CsPbI$_3$ nanocrystals with nuclear spin fluctuations'',
\href{https://doi.org/10.1103/s33c-m6hz}{Phys. Rev. B \textbf{113}, 035304 (2026)}.

\bibitem{abragam2012electron}
A.~Abragam, B.~Bleaney, \newblock \emph{Electron Paramagnetic Resonance of Transition Ions}, \newblock Oxford University Press, Oxford, (2012).


\bibitem{lovric2011hyperfine}
M. Lovri{\'c}, P. Glasenapp, D. Suter, B. Tumino, A. Ferrier, P. Goldner, M. Sabooni, L. Rippe, and S. Kr{\"o}ll,
``Hyperfine characterization and spin coherence lifetime extension in Pr$^{3+}$:La$_2$(WO$_4$)$_3$'',
\href{https://doi.org/10.1103/PhysRevB.84.104417}{Phys. Rev. B \textbf{84}, 104417 (2011)}.

\bibitem{lague2024spin}
G. Lague, F. Bernardot, V. Guilloux, L. Legrand, T. Barisien, J. S{\'a}nchez-Diaz, S. Galve-Lahoz, I. Sa{\"i}di, K. Boujdaria, J. P. Martinez-Pastor, et al.,
``Spin coherence and relaxation dynamics of localized electrons and holes in FAPbI$_3$ films'',
\href{https://doi.org/10.1021/acsphotonics.4c00632}{ACS Photonics \textbf{11}, 2770--2775 (2024)}.

\bibitem{kopteva2024weak}
N. E. Kopteva, D. R. Yakovlev, E. Kirstein, E. A. Zhukov, D. Kudlacik, I. V. Kalitukha, V. F. Sapega, O. Hordiichuk, D. N. Dirin, M. V. Kovalenko, et al.,
``Weak dispersion of exciton Land{\'e} factor with band gap energy in lead halide perovskites: Approximate compensation of the electron and hole dependences'',
\href{https://doi.org/10.1002/smll.202300935}{Small \textbf{20}, 2300935 (2024)}.

\bibitem{nestoklon2023tailoring}
M. O. Nestoklon, E. Kirstein, D. R. Yakovlev, E. A. Zhukov, M. M. Glazov, M. A. Semina, E. L. Ivchenko, E. V. Kolobkova, M. S. Kuznetsova, and M. Bayer,
``Tailoring the electron and hole Land{\'e} factors in lead halide perovskite nanocrystals by quantum confinement and halide exchange'',
\href{https://doi.org/10.1021/acs.nanolett.3c02349}{Nano Lett. \textbf{23}, 8218--8224 (2023)}.

\bibitem{meliakov2025lande}
S. R. Meliakov, E. A. Zhukov, V. V. Belykh, M. O. Nestoklon, E. V. Kolobkova, M. S. Kuznetsova, M. Bayer, and D. R. Yakovlev,
``Land{\'e} g-factors of electrons and holes strongly confined in CsPbI$_3$ perovskite nanocrystals in glass'',
\href{https://doi.org/10.1039/d4nr04602a}{Nanoscale \textbf{17}, 6522--6529 (2025)}.

\bibitem{turedi2022single}
B. Turedi, M. N. Lintangpradipto, O. J. Sandberg, A. Yazmaciyan, G. J. Matt, A. Y. Alsalloum, K. Almasabi, K. Sakhatskyi, S. Yakunin, X. Zheng, R. Naphade, S. Nematulloev, V. Yeddu, D. Baran, A. Armin, M. I. Saidaminov, M. V. Kovalenko, O. F. Mohammed, and O. M. Bakr,
``Single-crystal perovskite solar cells exhibit close to half a millimeter electron-diffusion length'',
\href{https://doi.org/10.1002/adma.202202390}{Adv. Mater. \textbf{34}, 2202390 (2022)}.

\bibitem{yang2022engineering}
C. Yang, J. Yin, H. Li, K. Almasabi, L. Guti{\'e}rrez-Arzaluz, I. Gereige, J.-L. Br{\'e}das, O. M. Bakr, and O. F. Mohammed,
``Engineering surface orientations for efficient and stable hybrid perovskite single-crystal solar cells'',
\href{https://doi.org/10.1021/acsenergylett.2c00431}{ACS Energy Lett. \textbf{7}, 1544--1552 (2022)}.

\end{thebibliography}

\begin{thebibliography}{38}%
\bibitem{belykh2022selectiveS}
V.~V.~Belykh and S.~R.~Melyakov,
``Selective measurement of the longitudinal electron spin
relaxation time $T_1$ of Ce$^{3+}$ ions in a YAG lattice:
Resonant spin inertia,''
\href{https://doi.org/10.1103/PhysRevB.105.205129}
{Phys. Rev. B \textbf{105}, 205129 (2022)}.

\bibitem{kirstein2022leadS}
E.~Kirstein, D.~R.~Yakovlev, M.~M.~Glazov, E.~Evers,
E.~A.~Zhukov, V.~V.~Belykh, N.~E.~Kopteva, D.~Kudlacik,
O.~Nazarenko, D.~N.~Dirin, M.~V.~Kovalenko, and M.~Bayer,
``Lead-dominated hyperfine interaction impacting the
carrier spin dynamics in halide perovskites,''
\href{https://doi.org/10.1002/adma.202105263}
{Adv. Mater. \textbf{34}, 2105263 (2022)}.
\end{thebibliography}
\end{document}